\documentclass[twocolumn,aps,pre,superscriptaddress,
letterpaper]{revtex4-1}
\usepackage{graphicx}
\usepackage[caption=false]{subfig}
\usepackage{sidecap}
\usepackage{amsmath}
\usepackage{amssymb}
\usepackage{color}
\usepackage{array}
\usepackage{placeins}
\usepackage[utf8]{inputenc}

\begin{document}


\title{
Towards a soft magnetoelastic twist actuator
}

\author{Lukas Fischer}
\email{lfischer@thphy.uni-duesseldorf.de}
\affiliation{Institut f{\"u}r Theoretische Physik II: Weiche Materie, 
Heinrich-Heine-Universit{\"a}t D{\"u}sseldorf, Universit{\"a}tsstra{\ss}e 1, D-40225 D{\"u}sseldorf, Germany}
\author{Andreas M. Menzel}
\email{menzel@thphy.uni-duesseldorf.de}
\affiliation{Institut f{\"u}r Theoretische Physik II: Weiche Materie, 
Heinrich-Heine-Universit{\"a}t D{\"u}sseldorf, Universit{\"a}tsstra{\ss}e 1, D-40225 D{\"u}sseldorf, Germany}

\date{\today}

\begin{abstract}
Soft actuators allow to transform external stimuli to mechanical deformations. Because of their deformational response to external magnetic fields, magnetic gels and elastomers represent ideal candidates for such tasks. Mostly, linear magnetostrictive deformations, that is, elongations or contractions along straight axes are discussed in this context. In contrast to that, we here suggest to realize a twist actuator that responds by torsional deformations around the axis of the applied magnetic field. For this purpose, we theoretically investigate the overall mechanical response of a basic model system containing discrete magnetizable particles in a soft elastic matrix. Two different types of discrete particle arrangements are used as starting conditions in the nonmagnetized state. These contain globally twisted anisotropic particle arrangements on the one hand, and groups of discrete helical-like particle structures positioned side by side on the other hand. Besides the resulting twist upon magnetization, we also evaluate different other modes of deformation. Our analysis supports the construction of magnetically orientable and actuatable torsional mixing devices in fluidic applications or other types of soft actuators that initiate relative rotations between different components. 
\end{abstract}

\maketitle

\section{Introduction}\label{Sec_Introduction}

Torsional actuators respond by a twist-type deformation to external stimuli. Most studies are concerned with linear actuators that contract or elongate along a certain axis upon actuation. However, there are several important prospective applications of twist actuators, for example microfluidic mixing, microscopic surgery tools, and prosthetics \cite{aziz2019torsional}. Depending on the application, a certain degree of softness of the actuator in combination with a certain degree of biocompatibility may be beneficial or even mandatory, particularly when it comes to medical applications. This is one of the reasons why so-called magnetic gels and elastomers (also commonly referred to as magnetorheological elastomers or ferrogels) \cite{filipcsei2007magnetic, jolly1996magnetoviscoelastic, odenbach2016microstructure, menzel2015tuned,schmauch2017chained, weeber2018polymer,weeber2019studying, stolbov2019magnetostriction, menzel2019mesoscopic, schumann2019microscopic} were introduced as important candidates for the construction of soft actuators \cite{zrinyi1996deformation,collin2003frozen,an2003actuating, filipcsei2007magnetic,zimmermann2007deformable,raikher2008shape, fuhrer2009crosslinking,bose2012soft,ilg2013stimuli,li2014state,maas2016experimental,lum2016shape, hines2017soft, becker2019magnetic}. These materials usually consist of magnetic or magnetizable colloidal particles embedded in an elastic, typically polymeric matrix. Such magnetic gels have the advantage that their distortions can be induced by external magnetic fields and the resulting deformation is typically reversible \cite{schumann2017situ}.

\begin{figure}
	\includegraphics[width=\linewidth]{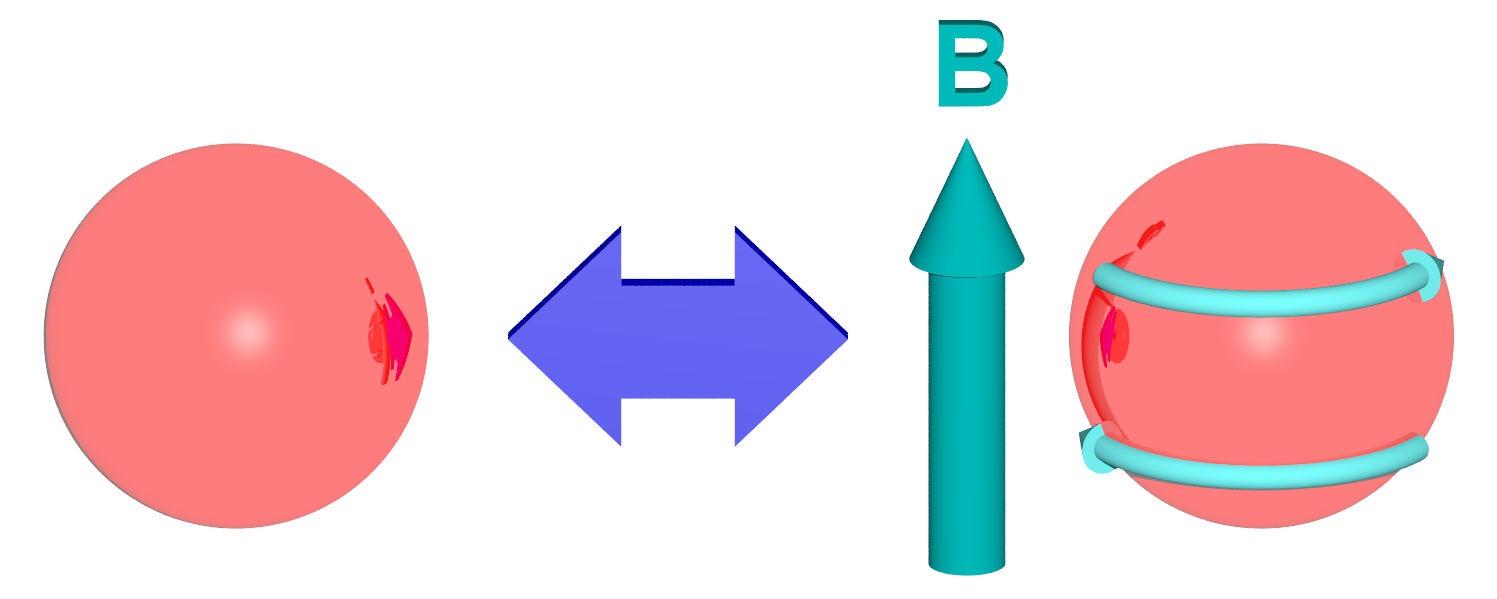}
	\caption{Illustration of the general idea and setup. The considered soft magnetoelastic composite system is spherical in overall shape. Upon application of a homogeneous external magnetic field $ \mathbf{B} $, it shows a reversible torsional twist deformation as indicated by the curved arrows on the right-hand side.}
	\label{fig_setup}
\end{figure}

To now generate magnetoelastic twist actuators in the form of magnetic gels or elastomers, see Fig.~\ref{fig_setup}, we suggest to build on the following previously explored insights. When the materials are fabricated in the presence of strong homogeneous external magnetic fields, chain-like structures of the inserted particles may form before the surrounding polymeric matrix is permanently established through corresponding chemical processes. Once the elastic matrix has reached its elastic solid state, these particle structures remain locked in the material, as can be seen in many experimental realizations \cite{collin2003frozen, coquelle2005magnetostriction,abramchuk2007novel,bose2007viscoelastic,chen2007microstructures,filipcsei2010magnetodeformation,gunther2011x,borbath2012xmuct, danas2012experiments}. 
One possible route to generate torsional actuators may be to additionally twist these chain-like aggregates, before the particle positions are fixed in the material by the final chemical crosslinking and establishing of the elastic polymeric matrix. This leads to self-supported torsional actuators. Such a concept is different from materials that are clamped at one end, contain anisotropic nonchiral structures, and are twisted by external magnetic fields that exert torques on the contained anisotropic aggregates \cite{monz2008magnetic, stanier2016fabrication}. Naturally, the situation that we consider is also different from studying how magnetic fields modify the stiffness of magnetic gels and elastomers when distorted by externally imposed torsional deformations \cite{abramchuk2007novel2, blom2012frequency, hashi2016dynamic, hoang2013development, lee2019design, merkulov2019research, sorokin2015hysteresis}. 

To realize soft torsional actuators, in our case, on the one hand, one may think of a globally, collectively twisted state of the whole set of embedded chain-like aggregates in the initial, cured state of the materials. On the other hand, one may consider each individual chain-like aggregate to show an initially twisted structure.

We start by considering globally twisted particle arrangements as initial states. To generate corresponding samples, a procedure of the following kind could be realistic. The approach is inspired by a protocol of synthesizing monodomain nematic liquid-crystalline elastomers \cite{kupfer1991nematic,kupfer1994liquid,ohm2010liquid}, consisting of liquid-crystal molecules that are chemically attached to or part of crosslinked polymeric networks \cite{urayama2007selected,krause2009nematic,ohm2010liquid}. Its scheme follows a two-step crosslinking process \cite{kupfer1991nematic,kupfer1994liquid,ohm2010liquid}, employing two crosslinkers of different speed of chemical reaction. The action of the first crosslinker generates a weakly crosslinked elastomeric sample that is stiff enough to already be uniaxially stretched. Maintaining this stretched state, in which the liquid-crystal molecules are on average uniaxially oriented in response to the imposed strain, the second crosslinker reacts and locks in this configuration. Along these lines, monodomain nematic samples, featuring an average uniaxial molecular liquid-crystalline alignment, are obtained. Such materials show pronounced nonlinear stress-strain properties when stretched perpendicular to the direction of nematic alignment \cite{kupfer1991nematic,kupfer1994liquid,urayama2007stretching,menzel2009response}. 

In our case of magnetic gels and elastomers, the two-step crosslinking process may be performed accordingly. First, under the presence of strong homogeneous external magnetic fields, uniaxially ordered chain-like aggregates of the magnetized particles form. They get locked into the sample by the generated surrounding elastic environment resulting from the quick action of the first crosslinker \cite{coquelle2005magnetostriction,abramchuk2007novel,bose2007viscoelastic,gunther2011x,borbath2012xmuct, danas2012experiments, zubarev2013effect, stepanov2008motion}. Finite gaps between the particles as considered below may result from previous coating of the particles, or by using surface-functionalized particles themselves as crosslinkers \cite{messing2011cobalt,barbucci2011novel,frickel2011magneto,ilg2013stimuli,weeber2015ferrogels,weeber2015ferrogels}. In a next step, this pre-crosslinked system is twisted around the anisotropy axis. This leads to a global twist of the contained chain-like particle aggregates. The sample is maintained in this state while the second, slower crosslinker is reacting chemically and establishing the final elastic matrix. In this way, the twisted structure gets permanently locked in. 

Another, possibly more academic procedure to generate example systems for investigations of the effects that we here predict might be to deposit the particles in a controlled way, maybe even by hand, at prescribed positions while generating the elastic environment layer by layer \cite{puljiz2016forces,puljiz2018reversible}. Even macroscopic spherical particles could be used for such proofs of concept \cite{chen2013numerical}. In this case, besides implementing globally twisted structures, one could also arrange the magnetizable particles in individual helices, positioned in an aligned way side by side. Maybe, in the future, such a deposition process can be automatized, as has recently been achieved for the production of magnetic microhelices \cite{ghosh2009controlled,peyer2013magnetic}. 

In the present work, we use such twisted discrete particle configurations as an input to calculate resulting magnetically induced overall deformations of corresponding elastic composite systems. Our theoretical approach is analytical, based on linear elasticity theory, and then evaluated numerically. To achieve such an analytical approach, we concentrate on elastic systems of overall spherical shape. The degree of initial twist is varied and the consequences of such variations are analyzed, both by numerical evaluations and by simplified analytical considerations. 
Both the globally twisted structures as well as several individual helical-like aggregates arranged side by side are addressed. 

We give a brief overview on our theoretical approach in Sec.~\ref{Sec_Framework}, together with a motivation of our chosen parameter values. After that, in Sec.~\ref{Sec_Global}, the torsional actuation of systems containing globally twisted particle configurations are addressed. In Sec.~\ref{Sec_Helix}, we consider particle arrangements of helical aggregates positioned side by side. To further facilitate the understanding, we compare the resulting twisting deformation to a minimal analytical consideration in Sec.~\ref{Sec_Force}. We conclude in Sec.~\ref{Sec_final}. 

\section{Theoretical framework}\label{Sec_Framework}

To perform the following evaluations, we build on our methods developed in Ref.~\onlinecite{JCP}. We assume that the elastic material used for the magnetorheological elastomer and containing the magnetizable particles is spatially isotropic as well as homogeneous. Moreover, we confine ourselves to small deformations (up to a couple of percent) so that we can use linear elasticity theory. This allows us to superimpose the deformations resulting from each internal force center.

Consequently, we describe this material via only two elastic coefficients, namely the shear modulus $ \mu $ and the Poisson ratio $ \nu $. They quantify the stiffness and compressibility of the material, respectively. A Poisson ratio $ \nu $ of $ 1/2 $, representing an upper bound \cite{landau1986theory}, describes incompressible materials. However, the Poisson ratio can reach negative values as well, down to $ -1 $ \cite{landau1986theory}. In these cases, the corresponding material is called auxetic, implying that when stretched along one axis it will show expansion to the lateral directions instead of contraction. 

Generally, the response of the elastic material to an applied force density $ \mathbf{f}(\mathbf{r}) $ inside it is then quantified by the so-called Navier--Cauchy equations \cite{cauchy1828exercises},
\begin{equation}\label{NC}
\mu\Delta \mathbf{u}(\mathbf{r}) + \frac{\mu}{1-2\nu}\nabla\nabla\cdot\mathbf{u}(\mathbf{r}) = {}-\mathbf{f}(\mathbf{r}),
\end{equation}
where $ \mathbf{u}(\mathbf{r}) $ denotes the displacement field at position $ \mathbf{r} $.
In our case, the elastic material forms a free-standing elastic sphere of radius $ R $. Fortunately, an analytical solution for Eq.~\eqref{NC} in this case is available in terms of the corresponding Green's function. $ \mathbf{f}(\mathbf{r}) $ then specifies the effect of a point-like force center inside the elastic sphere. We were able to transfer this solution to the case of a free-standing sphere of free surface \cite{JCP}, starting from previous work that considered the sphere embedded in an elastic background material \cite{walpole2002elastic}. This analytical solution for the elastic part of the problem was afterwards implemented numerically.

Next, to include the magnetic effects of magnetorheological gels and elastomers, we distributed magnetic inclusions at prescribed positions inside the elastic material, see Sec.~\ref{Sec_Global} and Sec.~\ref{Sec_Helix}. We always assume the magnetic inclusions to be sufficiently far apart from each other so that we can describe their magnetic signature as magnetic dipoles. As a further simplification, we assume that the magnetic dipolar moment  $\mathbf{m}=m\mathbf{\hat{m}}$, where $m=|\mathbf{m}|$, is identical for all inclusions. In experiments, such a situation could be realized by applying a strong external magnetic field that magnetizes all (identical) inclusions to saturation. 

In this case, the magnetic dipole--dipole forces are given by \cite{jackson1962classical}
\begin{equation}
\mathbf{F}_i= -\, \frac{3 \mu_0 m^2}{4\pi} \sum_{\substack{j=1 \\ j \neq i}}^{N} 
\frac{5 \mathbf{\hat{\bar{r}}}_{ij} \!\left( \mathbf{\hat{m}} \cdot \mathbf{\hat{\bar{r}}}_{ij}\right)^2 - \mathbf{\hat{\bar{r}}}_{ij} - 2 \mathbf{\hat{m}} \!\left( \mathbf{\hat{m}} \cdot \mathbf{\hat{\bar{r}}}_{ij} \right) }{\bar{r}_{ij}^4},
\label{eq_magn_dipole}
\end{equation}
where $ \mathbf{F}_i $ is the force exerted by all other inclusions on the $ i $th inclusion. Moreover, $\mu_0$ denotes the magnetic vacuum permeability, $\mathbf{\bar{r}}_i$ marks the position of the $i$th inclusion, the difference vector of positions is given by  $\mathbf{\bar{r}}_{ij}=\mathbf{\bar{r}}_i-\mathbf{\bar{r}}_j=\bar{r}_{ij}\mathbf{\hat{\bar{r}}}_{ij}$ with $\bar{r}_{ij}=|\mathbf{{\bar{r}}}_{ij}|$ ($i,j=1,...,N$), and $N$ sets the number of magnetized inclusions. The resulting force density inserted into Eq.~\eqref{NC} based on Eq.~\eqref{eq_magn_dipole} is
\begin{equation} \label{eq_f_density}
\mathbf{f}(\mathbf{r}) = \sum_{i=1}^{N} \mathbf{F}_i \, \delta(\mathbf{r}-\mathbf{\bar{r}}_i),
\end{equation}
where $ \delta(\mathbf{r}) $ represents the Dirac delta function and we thus assume point-like magnetic force centers.

After rescaling lengths by $ R $ and forces by $ \mu R^2 $, the strength of the magnetic forces relative to the elastic restoring forces is characterized by a dimensionless force coefficient $ 3 \mu_0 m^2/ 4\pi \mu R^6 $. Its value is set to $ 5.4 \times 10^{-8} $ for all that follows, as inspired by realistic experimental parameters \cite{JCP}. The inclusions are assumed to be of spherical shape as well, with their radius set to $ a = 0.02 R $. 

To include the effect of the induced elastic distortions on the positions of the magnetized inclusions and thus on the resulting magnetic forces and vice versa, an iterative scheme had been developed, see Ref.~\onlinecite{JCP}.
Finally, to characterize the induced overall deformations and capabilities of actuation, we evaluate the resulting displacement field on 49152 surface points of the elastic sphere. These points are approximately evenly distributed with positions generated by the HEALPix package (http://healpix.sourceforge.net) \cite{HEALPix}.

For the problem at hand, we choose the $ z $-axis to always coincide with the magnetization direction of the magnetic inclusions, i.e.\ $ \mathbf{\hat{m}} = \mathbf{\hat{z}} $. Moreover, we express the displacement of each surface point using spherical coordinates as
\begin{align}
\mathbf{u}\big(\mathbf{r}(\theta, \varphi)\big) =& \; u^{\bot}(\theta, \varphi) 
\begin{pmatrix}
\sin \theta \cos \varphi \\
\sin \theta \sin \varphi \\
\cos \theta
\end{pmatrix} \nonumber \\
&+\, u^{\theta}(\theta, \varphi) 
\begin{pmatrix}
\cos \theta \cos \varphi \\
\cos \theta \sin \varphi \nonumber \\
-\sin \theta
\end{pmatrix} \\
&+\, u^{\varphi}(\theta, \varphi) 
\begin{pmatrix}
-\sin \varphi \\
\cos \varphi \\
0
\end{pmatrix}
\end{align}
with
\begin{align}
\mathbf{r}(\theta, \varphi) =& \, R\begin{pmatrix}
\sin \theta \cos \varphi \\
\sin \theta \sin \varphi \\
\cos \theta
\end{pmatrix}.
\end{align}
Thus, the components $ u^{\bot} $, $ u^{\theta}$, and $ u^{\varphi} $ describe displacements inwards or outwards of the elastic surface, tangential deformations along the polar direction and tangential deformations along the azimuthal direction, respectively. Below, the latter coefficient $ u^{\varphi} $ will become particularly important to quantify the overall  twisting deformation.

To associate the resulting displacement field with different modes of overall deformation, we perform spherical harmonic expansions of $ u^{\bot} $, $ u^{\theta}$, and $ u^{\varphi} $. We use the same definitions for spherical harmonics, especially concerning the Condon--Shortley phase, as in Ref.~\onlinecite{jackson1962classical}. The most relevant spherical harmonics for our analysis are given by $Y_{00}=\sqrt{1/4\pi}$, $ Y_{10}= \sqrt{3/4\pi}\cos\theta $, and $Y_{20}=\sqrt{5/16\pi}\left(3\cos^2\!\theta-1\right)$.

As announced above, we then focus on the resulting overall torsional deformations for two types of spatial arrangements of the magnetizable inclusions: globally twisted and side-by-side aligned helical structures, see Secs.~\ref{Sec_Global} and \ref{Sec_Helix}, respectively. The degree of initial structural twist in the nonmagnetized state is quantified by a parameter $ \gamma $, see below for its definition. In both cases, we confine the initial positions of the inclusions by requiring a minimal distance of $ 3a = 0.06 R $ to the elastic spherical surface.

\section{Globally twisted structures} \label{Sec_Global}
\begin{figure}
\includegraphics[width=\linewidth]{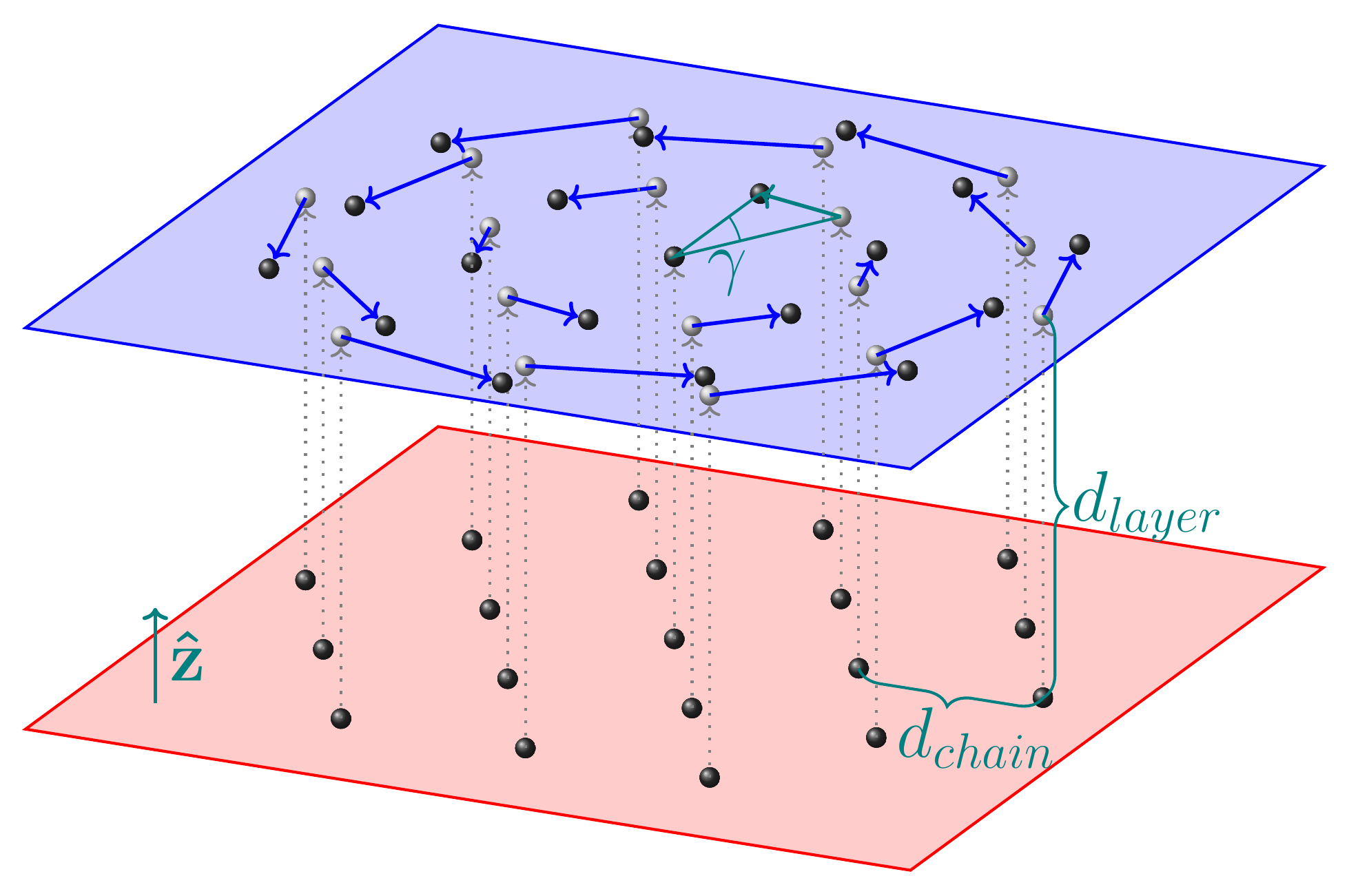}
\caption{Illustration of two layers of hexagonally arranged magnetizable inclusions inside the elastic material. $ d_{layer} $ sets the vertical distance between two layers, $ d_{chain} $ the in-plane particle distance. We set $ d_{chain} = 0.25R > d_{layer} = 0.11 R$, which implies vertically aligned chain-like aggregates.  Here, for illustration, $ d_{layer} $ is exaggerated. The upper arrangement shows a rotation by an angle $ \gamma $ relative to the lower arrangement, where we chose $ \gamma = \pi / 6 $ for reasons of visibility. To emphasize the twist from layer to layer, we plot the positions corresponding to the lower layer in the upper layer as gray spheres, together with a dotted arrow that shows their vertical identification. Having applied a rotation by $ \gamma $ to the structure from the lower layer (gray), the positions marked by dark spheres result. We indicate this in-plane rotational displacement by blue in-plane arrows. In the teal triangle, we illustrate the definition of the angle $ \gamma $.}
\label{fig_hexagonal}
\end{figure}

\begin{figure}
\includegraphics[width=\linewidth]{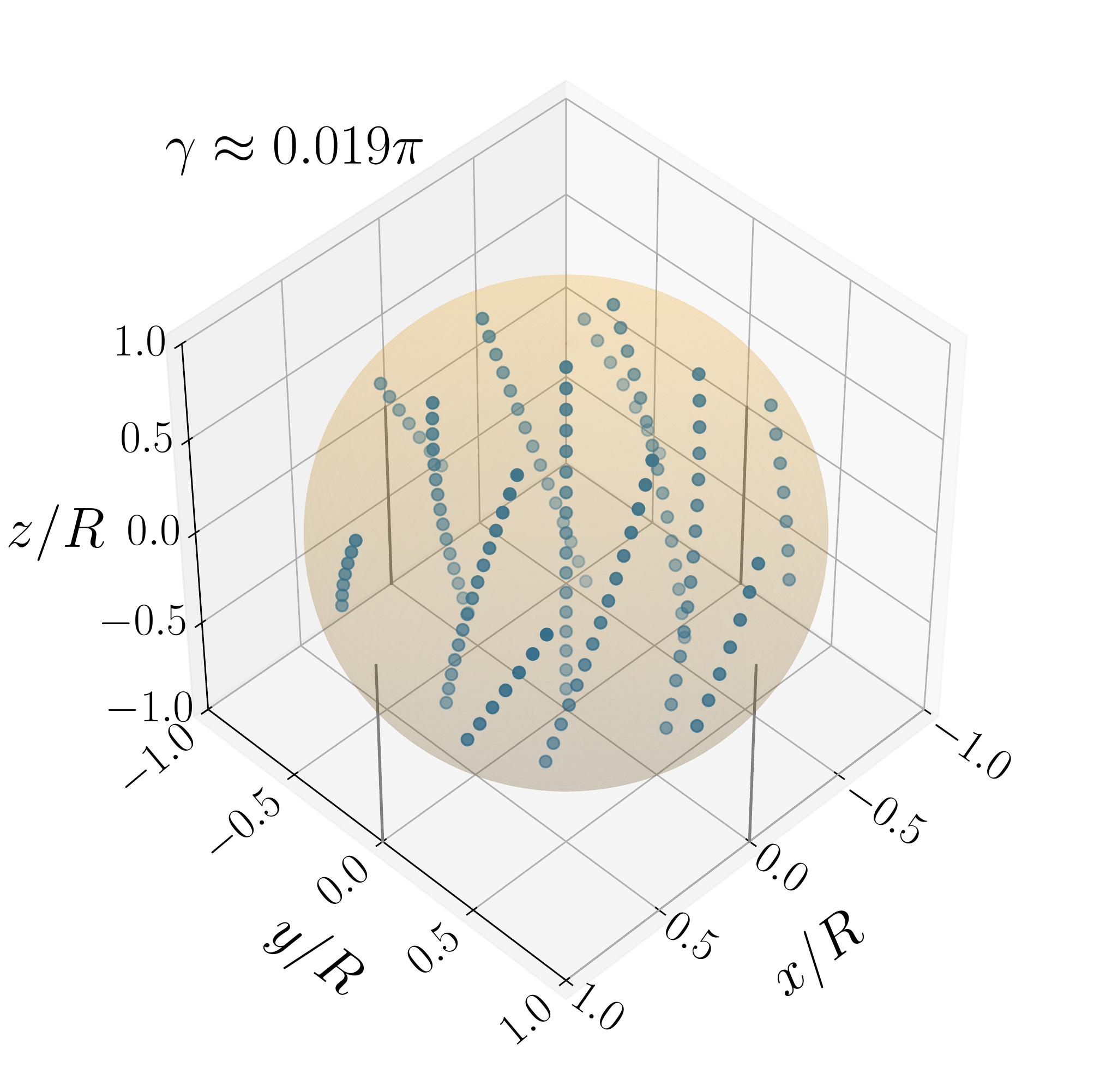}
\caption{Illustration of an example for the initial structure of the magnetizable inclusions, indicated as small spheres, inside the larger elastic sphere. This structure is generated from hexagonally arranged parallel chain-like aggregates of particles, where each horizontal layer of particles is rotated relative to the next particle layer underneath by an angle $ \gamma $, see Fig.~\ref{fig_hexagonal}. In this illustration, we chose $ \gamma \approx 0.019 \pi $. Moreover, for better visibility, we here set $ d_{chain} = 0.5 R $. Instead, for our actual numerical evaluation, we used a value of $ d_{chain} = 0.25 R $.}
\label{fig_setup_global}
\end{figure}

To numerically generate the globally twisted structures, we start from layers of hexagonally arranged magnetic inclusions \cite{stepanov2008motion, ivaneyko2012effects, metsch2016numerical, zahn1999two}. These layers are all oriented parallel to the $ xy $-plane and spaced equally from each other in their normal direction by a distance $ d_{layer} = 0.11 R $. The center layer is located in the plane $ z=0 $. In the initial, nonmagnetized situation, the hexagonal particle arrangements within each layer are in a state rotated by an angle of $ \gamma z/ d_{layer} $ relative to the arrangement in the plane $ z=0 $. This corresponds to a globally twisted configuration of the inclusions when compared to straight chain-like aggregates aligned parallel to the $ z $-axis. Here, we consider small angles $ \gamma \lesssim 0.159 \pi $ to preserve the chain-like structure, see  Fig.~\ref{fig_hexagonal}. The lattice constant within each plane, which equals the lateral distance between the chains, is set to $ d_{chain} = 0.25 R $. Overall, this leads to 623 magnetizable inclusions in 55 chains. An illustration of an initial structure is presented in Fig.~\ref{fig_setup_global}, where we have, however, increased $ d_{chain} $ to $ 0.5 R$ for clarity.
In the numerical evaluation, we consider the range $ 0 \lesssim \gamma \lesssim 0.159 \pi$ in steps of approximately $ 0.0016 \pi $. We distinguish four possible values of the Poisson ratio: $ \nu = 0.5 $ (incompressible), $ \nu = 0.3 $, $ \nu = 0 $, and $ \nu = -0.5 $ (auxetic).


As a first step, we focus on the following spherical harmonic expansion parameters for the resulting overall surface distortions: $ u^{\bot}_{00}$, $ u^{\bot}_{20}$, and $ u^{\varphi}_{10} $. The coefficient $ u^{\bot}_{00} $ quantifies overall changes in volume of the composite material. Positive values correspond to an increase in volume, while negative values correspond to a decrease in volume. Next, the coefficient $ u^{\bot}_{20}$ describes a relative elongation ($ u^{\bot}_{20} > 0 $) or contraction ($ u^{\bot}_{20} < 0$) along the direction of magnetization, here along the $ z $-axis. Most important for our investigation in the present context is the parameter $ u^{\varphi}_{10} $. This coefficient is set by the lowest mode of a twist-type deformation around the $z$-axis. For a counter-clockwise rotation of the upper hemisphere against the lower hemisphere, it becomes $ u^{\varphi}_{10} > 0 $. For a reversed mutual sense of rotation, one obtains $ u^{\varphi}_{10} < 0 $. 

\begin{figure}
	\centering
	\includegraphics[width=7.64468cm]{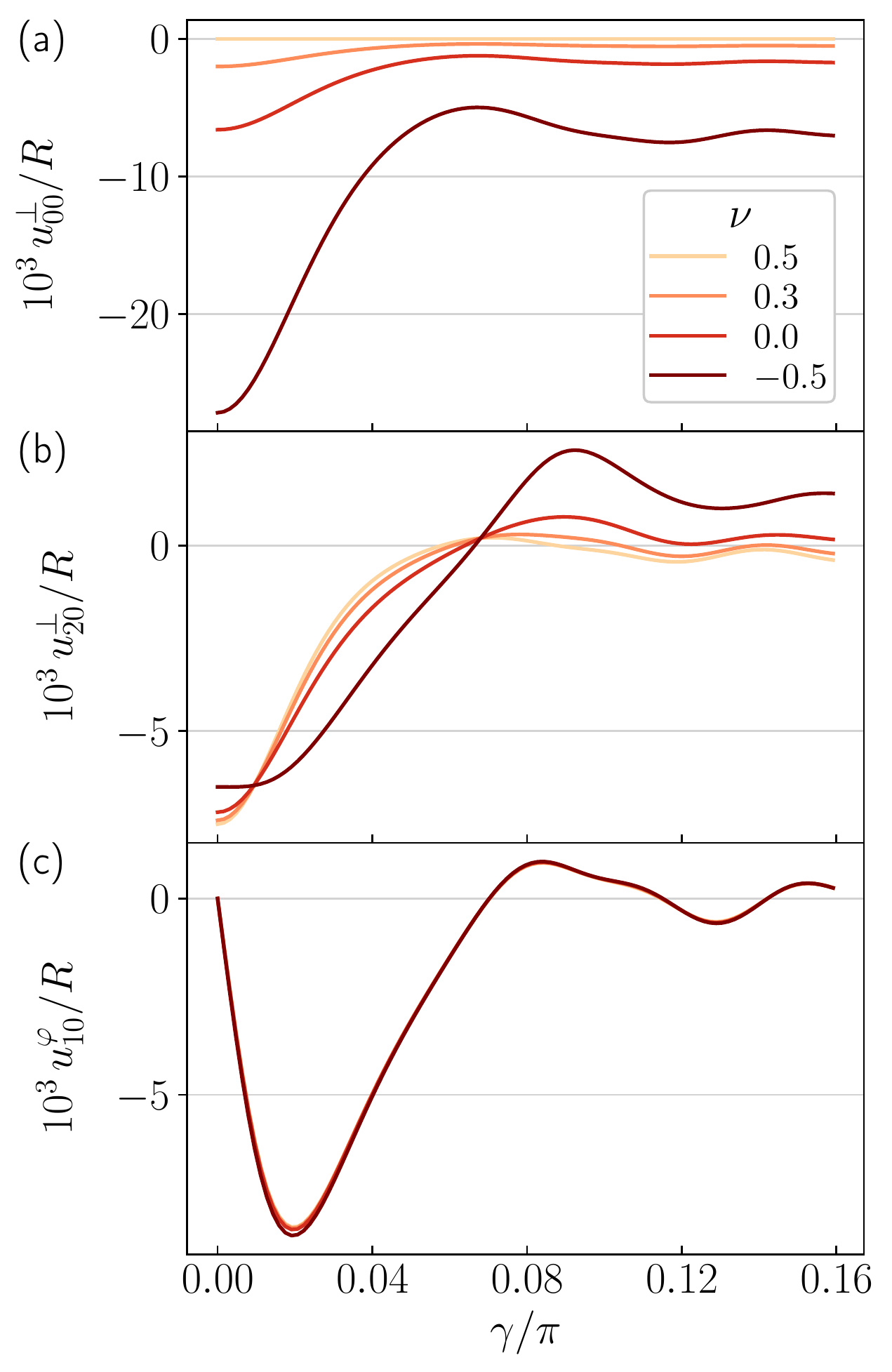}
	\caption{Resulting overall surface displacement field of the spherical magnetoelastic composite upon magnetization for an initially globally twisted configuration of the magnetizable inclusions. To quantify perpendicular surface displacements, we plot the two coefficients (a) $ u^{\bot}_{00}$ and (b) $ u^{\bot}_{20}$, indicating overall volume changes and overall elongation along the magnetization direction relative to lateral contraction, respectively. To quantify the lowest mode of an overall twist deformation around the magnetization axis, we plot the coefficient (c) $ u^{\varphi}_{10} $. In all three cases, we display the behavior with increasing angle $\gamma$, characterizing the global twist of the initial nonmagnetized structure of inclusions (see Fig.~\ref{fig_hexagonal} for the definition of $\gamma$). Moreover, we show graphs for the four different values of the Poisson ratio, namely $ \nu=0.5 $, $ \nu=0.3 $, $ \nu=0 $, and $ \nu=-0.5 $.}
	\label{fig_global}
\end{figure}

The three coefficients $ u^{\bot}_{00} $, $ u^{\bot}_{20}$, and $ u^{\varphi}_{10} $ are shown in Fig.~\ref{fig_global} when the aforementioned particle structures are magnetized.
We have not included the data for negative values of $ \gamma $ because the curves in Figs.~\ref{fig_global}(a) and \ref{fig_global}(b) are mirror symmetric with respect to the line $ \gamma=0 $, while the curve in  Fig.~\ref{fig_global}(c) features a point symmetry with respect to the origin.

As a first result, we infer from Fig.~\ref{fig_global}(a) that the overall volume is constant ($ u^{\bot}_{00} \approx 0$) for $ \nu = 0.5 $, as expected for an incompressible material. With decreasing Poisson ratio, we find that the elastic sphere shrinks more and more upon magnetization. Naturally, this volume decrease is maximal for $ \gamma = 0 $, i.e.\ straight chains of magnetizable inclusions. In this case, the induced attraction between the particles along each chain is strongest. When increasing $ \gamma $, the volume decrease becomes smaller and oscillates for higher values of $ \gamma $.

Similarly, we infer from Fig.~\ref{fig_global}(b) that the overall contraction along the magnetization direction relative to a lateral expansion, as quantified by $ u^{\bot}_{20} $, is strongest for $ \gamma = 0 $ for the same reason as above. This effect is most pronounced for incompressible materials because the contraction along the field implies lateral expansions for reasons of volume conservation. In contrast to that, the auxetic nature for $ \nu = -0.5 $ counteracts the lateral expansion for $ \gamma = 0 $.  The oscillations for increasing values of $ \gamma $ can be found in this coefficient as well.

When we focus on the behavior of $ u^{\varphi}_{10} $ in Fig.~\ref{fig_global}(c), we observe that it is almost independent of the Poisson ratio. This is expected because a pure twist-type deformation leaves the total volume unchanged. A small effect of the Poisson ratio is still present and can most likely be attributed to nonlinear effects revealed by our iterative scheme, i.e.\ to the effects of the resulting displacements of the magnetic inclusions which are larger for more compressible materials.
Furthermore, we do not observe any torsional deformation for $ \gamma = 0 $ because our initial  configuration is not twisted in this case. Increasing $ \gamma $ from zero, we see that the corresponding values of $ u^{\varphi}_{10} $ first become more and more negative. The sign here represents the sense of the induced torsional deformation of the composite which is opposing the sense of initial twist of the initial structure. We reach a maximum magnitude of this twist deformation at $ \gamma \approx 0.019 \pi $. For larger values of $ \gamma $, the magnitude of $ u^{\varphi}_{10} $ again decreases. This effect results from the increasing distance between the inclusions with increasing $ \gamma $ implying a decreasing magnetic interaction. At even larger values of $ \gamma $, $ u^{\varphi}_{10} $ oscillates around zero. We return to this feature in Sec.~\ref{Sec_Force}.

In practice, one would typically be interested in the situation of maximum observed effect. We therefore concentrate on the system for $ \gamma \approx 0.019 \pi $. First, we checked how the magnitude of the induced torsion around the $z$- axis varies with the height $z$ above or below the horizontal center plane (the $xy$-plane). For this purpose, we
calculated the average azimuthal angular displacement of the horizontal plane parallel to the $ xy $-plane at height $ z $ as 
\begin{equation}
	\Delta \varphi(z) = \langle\arctan\left(\frac{ u^{\varphi}} {\sqrt{R^2-z^2}}\right)\rangle_{z},
\end{equation}
where $ \langle \dots \rangle_{z} $ denotes an average over all surface points at which $ u^{\varphi} $ was evaluated at a given height $ z $. We found that this quantity is approximately proportional to $ z $.  Furthermore, we find that it is nearly independent of the Poisson ratio, in agreement with the behavior of $u^{\varphi}_{10}$ in Fig.~\ref{fig_global}(c).

\begin{figure}
	\includegraphics[width=\linewidth]{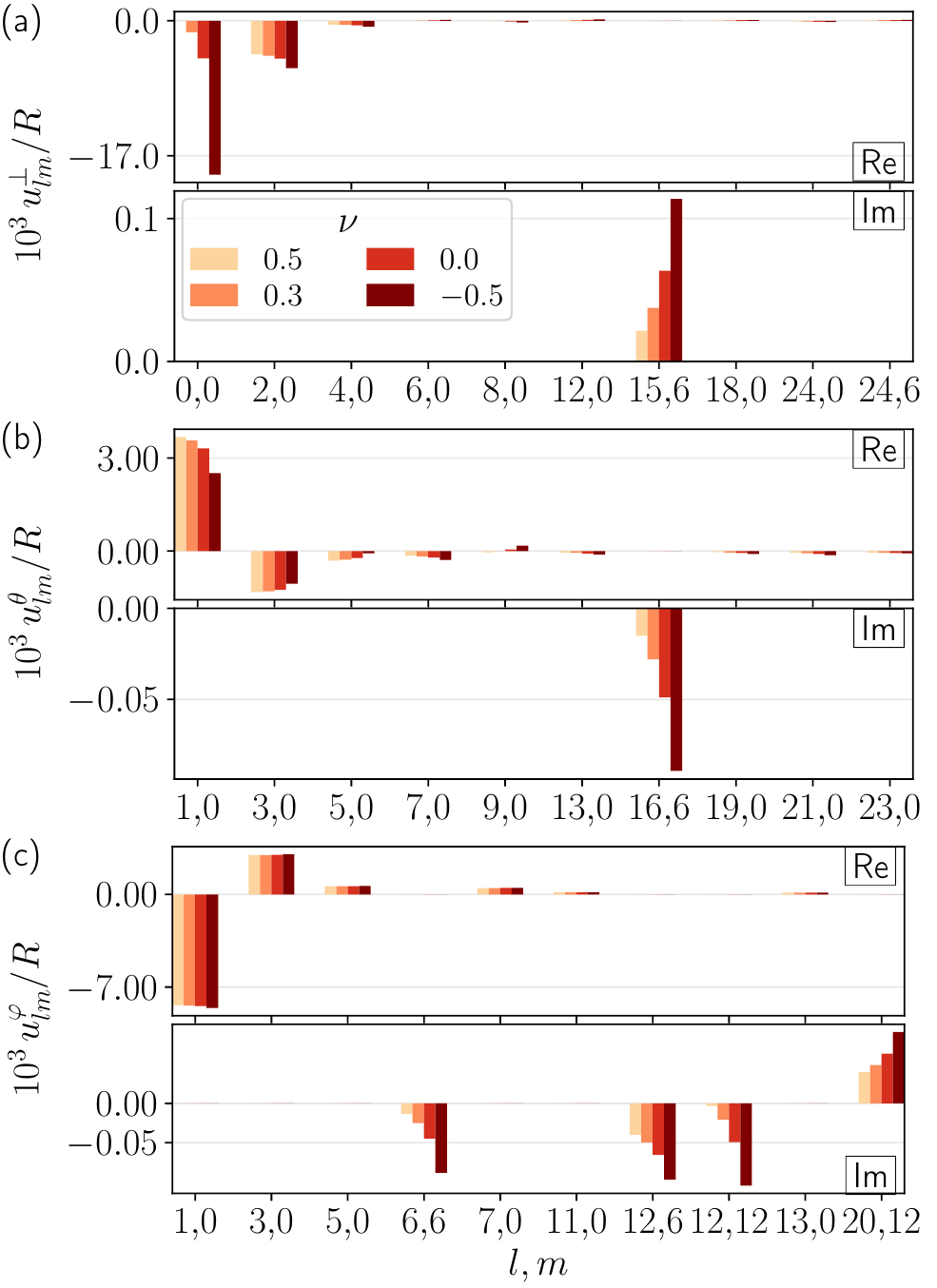}
	\caption{For the same systems considered in Fig.~\ref{fig_global}, we depict for $ \gamma \approx 0.019 \pi $ the values of the expansion coefficients into spherical harmonics for the three components (a) $ u^{\bot} $, (b) $ u^{\theta} $, and (c) $ u^{\varphi} $ of the overall surface displacement field. The value of $ \gamma $ is the same as in Fig.~\ref{fig_setup_global}. The real part is always plotted in the top line, while the lower line illustrates the imaginary part of the corresponding spherical harmonic expansion coefficient. We use bar plots with the four colors corresponding to the four possible values of the Poisson ratio $ \nu=0.5 $, $ \nu=0.3 $, $ \nu=0 $, and $ \nu=-0.5 $. In this way, the values of the expansion coefficients for ten representative modes are displayed for the three components of the surface displacement field. Particularly, we note the dominating character of the mode $ (l,m) = (1,0) $ for $ u^{\varphi} $, which is associated with the type of twist actuation upon magnetization that we here focus on.}
	\label{fig_global_spectrum}
\end{figure}

Next, in Fig.~\ref{fig_global_spectrum}, we provide additional information on the importance of different modes involved in the overall surface displacement, obtained by our expansion of the perpendicular and tangential components of the surface displacement field into spherical harmonics. Again, we concentrate on the value of $ \gamma \approx 0.019 \pi $ and we use the same four values of the Poisson ratio as in Fig.~\ref{fig_global}. 
We select the expansion coefficients $ a_{lm} $ of ten representative spherical harmonic modes for each component of the displacement field according to the following scheme. First, for each mode the value of $ a_{lm} $ of highest magnitude is identified from the four values for the different Poisson ratios $ \nu $. These $ a_{lm} $ are then ordered according to their absolute values, and we find the labels $ l,m $ for the ten largest ones. Due to the high degree of symmetry of our configurations, the most dominant modes are those of $ m=0 $. However, we observe nonvanishing modes that depend on $ \varphi $ as well, characterized by $ m \neq 0 $. This leads to complex expansion coefficients. Since $ u^{\bot} $, $ u^{\theta}$, and $ u^{\varphi} $ are real, we can find values for negative $ m $ via the relation $ a_{l(-m)} = (-1)^{m} a_{lm}^*$, where the star denotes complex conjugation. Consequently, for real $ a_{lm} $ the corresponding spherical harmonics result together with $ a_{l(-m)} $ in a $ \cos \left(m\varphi\right)  $ mode, while purely imaginary $ a_{lm} $ result in a $ -\!\sin \left(m\varphi\right) $ mode. The real and the imaginary part of $ a_{lm} $ are shown separately in the plots.

Figure \ref{fig_global_spectrum} confirms that those coefficients that we have been concentrating on so far indeed dominate the spectrum. For $ u^{\bot}$, see Fig.~\ref{fig_global_spectrum}(a), these correspond to an overall volume change ($ l=m=0 $), especially for auxetic materials and except for $ \nu = 0.5 $, and to an overall contraction along the magnetization direction relative to a lateral expansion ($ l=2,\,m=0 $), with small higher-order corrections. All coefficients odd in $ l $ for $ m=0 $ are approximately zero here. We observe some very small contributions related to the six-fold symmetry about the $ z $-axis in the modes of $ l=15,\,m=6 $ and $ l=24, \, m=6 $.

Turning to $ u^{\theta} $ in Fig.~\ref{fig_global_spectrum}(b), significantly smaller absolute values of the expansion coefficients are obtained. Here, as for $ u^{\varphi} $ in Fig.~\ref{fig_global_spectrum}(c), the coefficients even in $ l $ vanish approximately for $ m=0 $, in contrast to the case for $ u^{\bot} $. The most important contribution to $ u^{\theta} $ in the mode $ l=1, \, m=0 $ corresponds to an overall surface displacement towards the equator on both the upper hemisphere and the lower hemisphere upon magnetization. In the incompressible case, this effect is most pronounced as we then have the strongest expansion of the sphere in the lateral directions. Again, higher-order contributions emerge which strengthen the aforementioned effect in the vicinity of the equatorial plane.

Considering $ u^{\varphi} $ in Fig.~\ref{fig_global_spectrum}(c) reveals the most important mode in the present context, associated with the twist deformation through magnetization. As noted already above, the mode $ l=1, \, m=0 $ is associated with a rotation around the magnetization direction of the upper hemisphere relative to the lower hemisphere. This mode dominates the overall behavior by its absolute value [only exceeded by the mode corresponding to overall volume expansion for the auxetic case $\nu=-0.5$ in Fig.~\ref{fig_global_spectrum}(a)]. Near the equatorial plane, higher-order modes in combination still support the effect of the upper hemisphere being rotated relatively to the lower hemisphere.

\FloatBarrier
\section{Helical structures} \label{Sec_Helix}

\begin{figure}
\includegraphics[width=\linewidth]{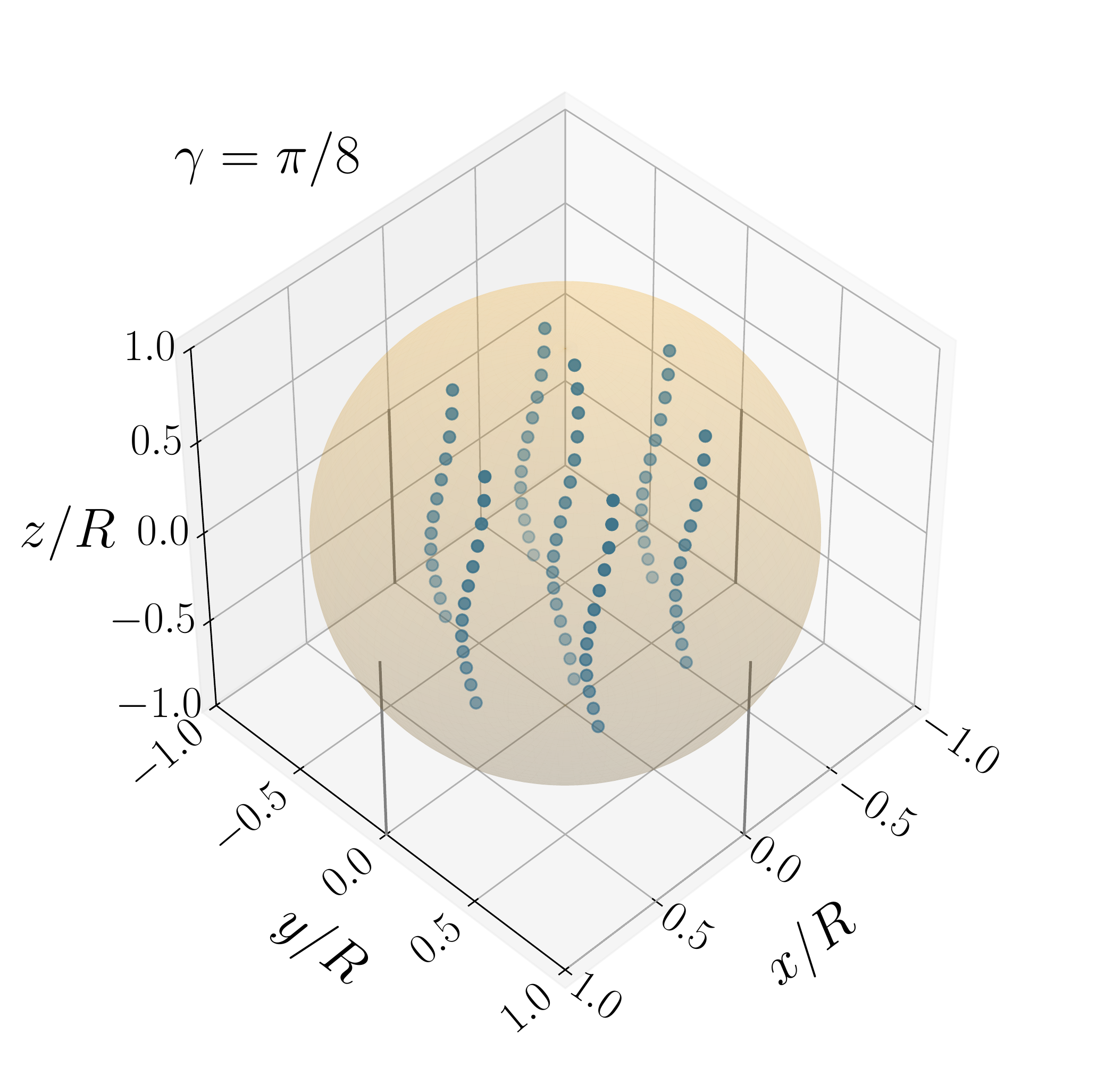}
\caption{Illustration of initially nonmagnetized particle structures composed of helical elements of magnetizable inclusions, arranged side by side. Within each layer parallel to the $xy$-plane, the particles form a hexagonal lattice of lattice constant $ d_{chain} = 0.5 R $. The layers have a vertical spacing of $ d_{layer} = 0.11 R $, see Eq.~\eqref{eq_helix_vec}. Furthermore, we here chose the radius of each helix to be $ r_{helix} = 0.05R $. In the depicted case, we set $ \gamma= \pi /8$.}
\label{fig_setup_helix1}
\end{figure}

\begin{figure}
\includegraphics[width=\linewidth]{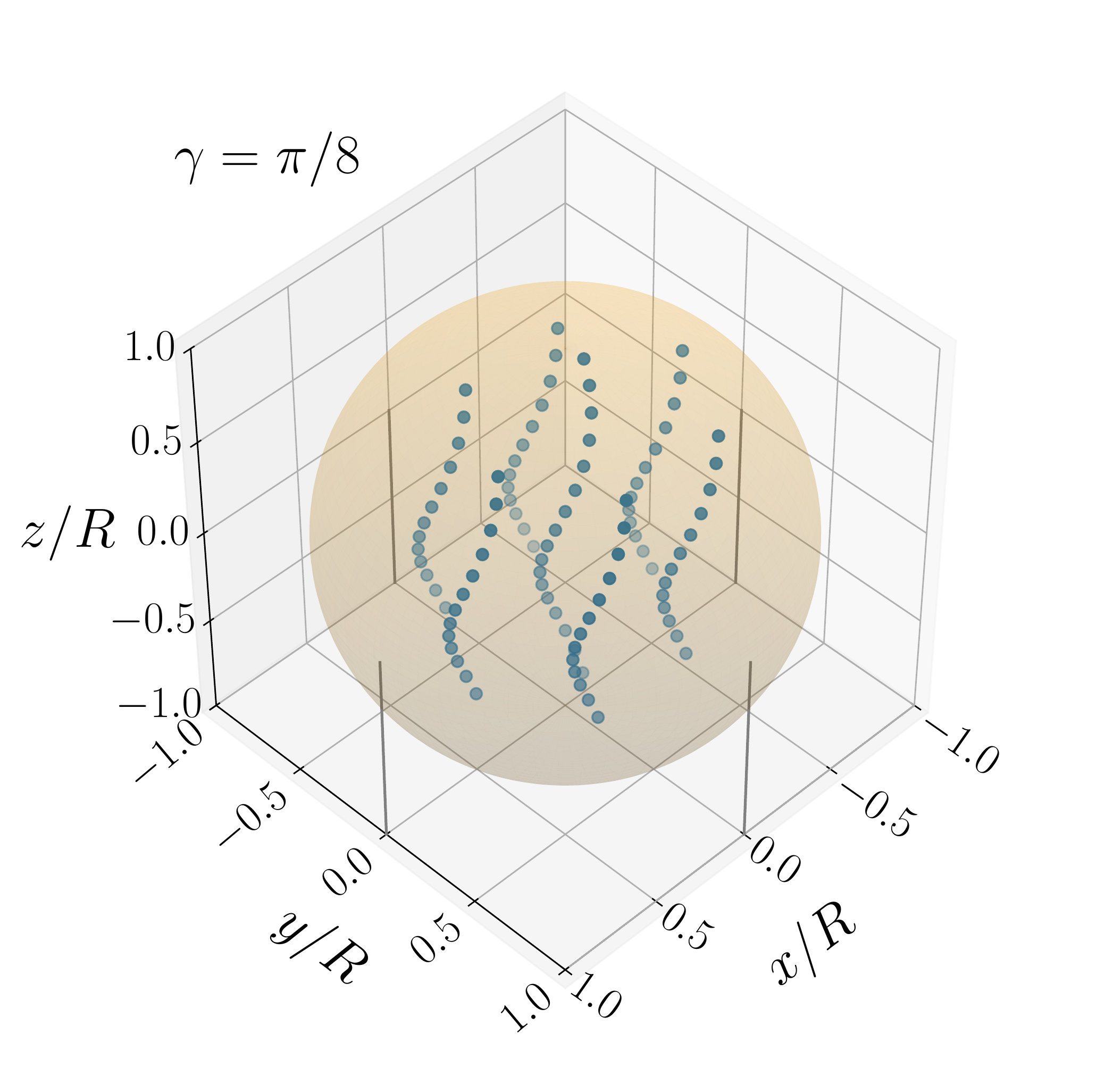}
\caption{Same as in Fig.~\ref{fig_setup_helix1}, but for $ r_{helix} = 0.1R $.}
\label{fig_setup_helix2}
\end{figure}

As a next step, we address helical structures of the magnetizable particles embedded in the same elastic spheres as before, arranged side by side. In contrast to the globally twisted structure  of parallel chain-like aggregates investigated in Sec.~\ref{Sec_Global}, we now consider each chain-like element by itself to feature an initial helical shape. To set up our numerical systems, we again start from hexagonal arrangements of aligned chain-like aggregates as before, this time for $ d_{chain} = 0.5 R $, i.e.\ for double the distance to each other. As above, the vertical distance of the horizontal layers of particles is set to $ d_{layer} = 0.11 R $. However, instead of initiating each layer rigidly rotated relatively to its upper and lower neighboring one, we now rigidly displace each layer laterally by adding a vector
\begin{align} \label{eq_helix_vec}
	\mathbf{r}_{helix}(z) &= r_{helix} \begin{pmatrix}
	\cos(\gamma z/ d_{layer})\\
	\sin(\gamma z/ d_{layer})\\
	0
	\end{pmatrix}.
\end{align}
This lateral shift introduces an additional parameter, namely $ r_{helix} $. Here, we show results for structures corresponding to two different values $ r_{helix} = 0.05R $ and $ r_{helix} = 0.1 R $, see Figs.~\ref{fig_setup_helix1} and \ref{fig_setup_helix2}, respectively. In both cases, we fit 95 magnetizable inclusions into our elastic sphere, avoiding inclusions that would need to be deleted for particular values of $ \gamma $.
Importantly, the overall structure in each case is no longer six-fold rotationally symmetric about the $ z $-axis nor globally screw-symmetric within the spherical boundaries.
In the center layer for $z=0$, all helices start with a particle deflection in the $ x $-direction, $ \mathbf{r}_{helix}(0) = r_{helix} \mathbf{\hat{x}} $, according to Eq.~\eqref{eq_helix_vec}. The resulting structures composed of helical aggregates are depicted in Figs.~\ref{fig_setup_helix1} and \ref{fig_setup_helix2}. Using our numerical approach, we evaluate the full range $ 0 \leq \gamma \leq 2\pi $ in steps of $ \pi / 360$.

\begin{figure}
	\centering
	\includegraphics[width=7.64468cm]{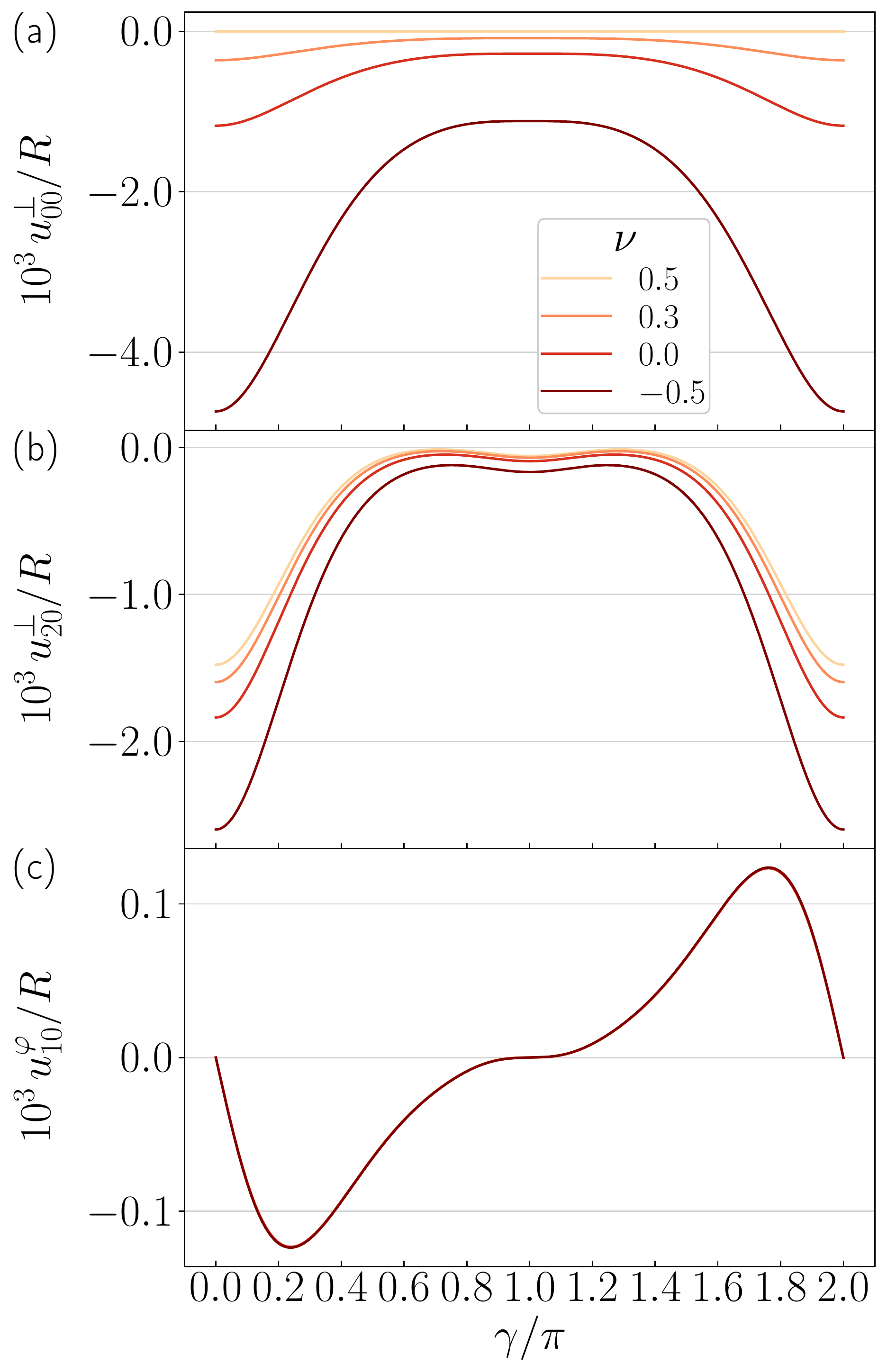}
	\caption{Same as in Fig.~\ref{fig_global}, but for the systems composed of helical structure elements of magnetizable inclusions arranged side by side instead of a globally twisted structure. Here, $ r_{helix} = 0.05R $, as in Fig.~\ref{fig_setup_helix1}.}
	\label{fig_helix1}
\end{figure}

\begin{figure}
	\centering
	\includegraphics[width=7.64468cm]{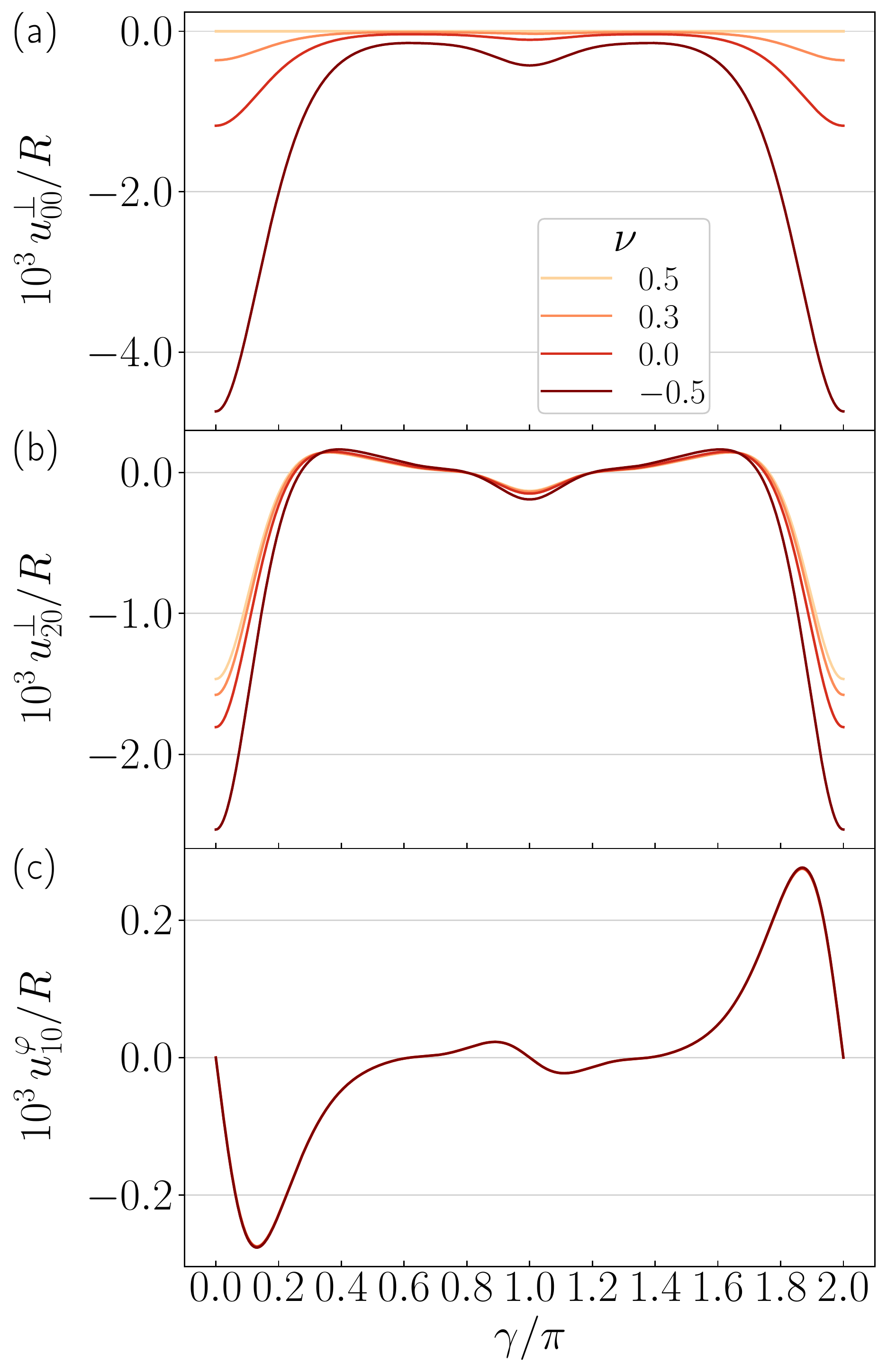}
	\caption{Same as in Fig.~\ref{fig_helix1}, but for setups with $ r_{helix} = 0.1R $.}
	\label{fig_helix2}
\end{figure}

As in Sec.~\ref{Sec_Global}, we first address the expansion coefficients $ u^{\bot}_{00}$, $ u^{\bot}_{20}$, and $ u^{\varphi}_{10} $ for the overall displacements upon magnetization as functions of $ \gamma $.
The curves in Figs.~\ref{fig_helix1}(a), \ref{fig_helix1}(b), \ref{fig_helix2}(a), and \ref{fig_helix2}(b) show a mirror symmetry with respect to the vertical line $ \gamma=\pi $, while those in Figs.~\ref{fig_helix1}(c) and \ref{fig_helix2}(c) feature a point symmetry with respect to the point $ \left(\gamma,u^{\varphi}_{10}\right) = (\pi,0) $. This is expected because $ Y_{00} $ and $ Y_{20} $ are even in $ z $, while $ Y_{10} $ is odd. Obviously, the results for helices of an initial twist $ \pi < \gamma < 2\pi $ can be mapped onto those for a corresponding initial twist of $ 2\pi - \gamma $. Illustratively, this corresponds to helices that only differ by their sense of twist.
The resulting displacements are of much smaller magnitude when compared to the results for the globally twisted arrangements in Fig.~\ref{fig_global}, which can already be expected from the lower total number of inclusions for the helical structures (95 here versus 623 inclusions in Fig.~\ref{fig_global}). 


We start by considering the configurations of $ r_{helix} = 0.05R $. In Fig.~\ref{fig_helix1}(a), we again find that the elastic sphere, except for $ \nu=0.5 $, shrinks as a whole, specifically for the smallest and largest values of $ \gamma $. For these values, the chains are straightest and therefore show the maximal internal longitudinal attractive forces. Furthermore, the absolute magnitude of overall contraction strongly increases with decreasing Poisson ratio, i.e.\ for more compressible spheres.

Next, we address in Fig.~\ref{fig_helix1}(b) the elongation along the magnetization relative to the lateral contraction. Qualitatively, we infer a similar behavior as in Fig.~\ref{fig_helix1}(a). Here, we observe a further, much smaller minimum for $ \gamma=\pi $ because we have effectively generated two chains of distance $ 2r_{helix} = 0.1 R $ out of each helix. Apart from that, auxetic materials show stronger relative contractions along the magnetization axis. 

Concerning the magnitude of the twist actuation quantified by Fig.~\ref{fig_helix1}(c), we again find a pronounced minimum, here around $ \gamma \approx 0.24 \pi $. In line with the point symmetry of the curve mentioned above, the corresponding maximum is located at $ \gamma \approx 1.76 \pi $. As in Sec.~\ref{Sec_Global}, $ u^{\varphi}_{10} $ as a measure for the twist actuation is approximately independent of the Poisson ratio. This behavior will also be discussed in Sec.~\ref{Sec_Force}.

Figure \ref{fig_helix2} shows corresponding results for $ r_{helix} = 0.1R $. The qualitative picture is similar to Fig.~\ref{fig_helix1}, with the same symmetries of the curves. We notice that the aforementioned minimum at $ \gamma = \pi $ is more pronounced and can be found in $ u^{\bot}_{00}$ [Fig.~\ref{fig_helix2}(a)] as well. Concerning the coefficient $u^{\varphi}_{10}$ quantifying the twist actuation, we see that the minimum is shifted to smaller values of $ \gamma $, namely to $ \gamma \approx 0.13 \pi $, and is increased in magnitude by a factor of approximately $ 2.23 $. Moreover, some oscillations together with positive values of $ u^{\varphi}_{10} $ occur at higher values of $ \gamma < \pi $.
Again, we will return to this topic in Sec.~\ref{Sec_Force}.

\begin{figure}
	\includegraphics[width=\linewidth]{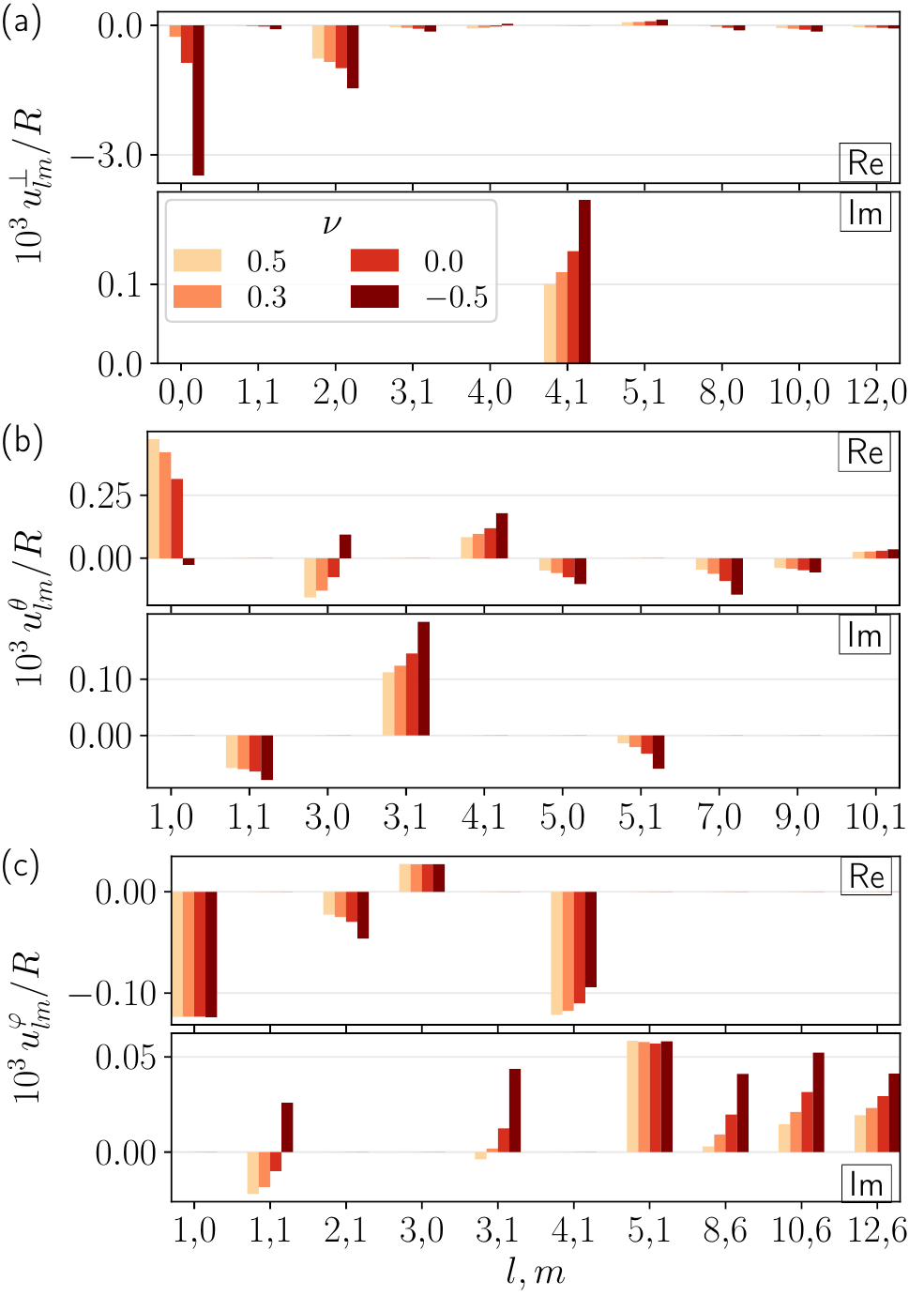}
	\caption{Same as in Fig.~\ref{fig_global_spectrum}, but here for a system of helical structures of $r_{helix}=0.05R$ arranged side by side, instead of a globally twisted structure, for $ \gamma \approx 0.24 \pi $. This value of $\gamma$ corresponds to the minimum of the curve in Fig.~\ref{fig_helix1}(c), i.e.\ the value of maximal twist actuation.}
	\label{fig_helix1_spectrum}
\end{figure}

\begin{figure}
	\includegraphics[width=\linewidth]{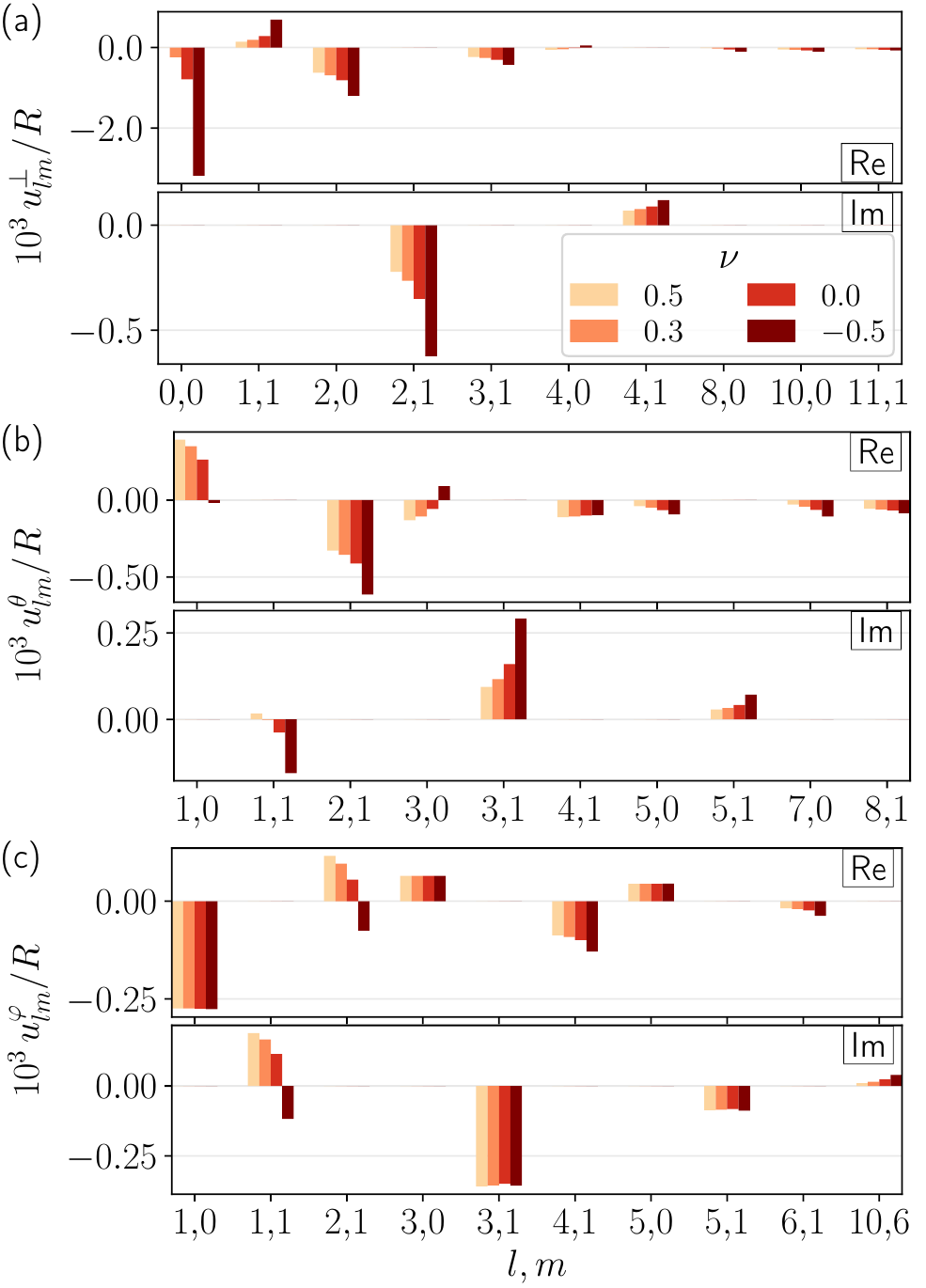}
	\caption{Same as in Fig.~\ref{fig_helix1_spectrum}, but for a setup with $ r_{helix} = 0.1R $ and for $ \gamma \approx 0.13 \pi $. The latter value identifies the minimum on the curve of Fig.~\ref{fig_helix2}(c).}
	\label{fig_helix2_spectrum}
\end{figure}

We continue with a discussion on the coefficients obtained from an expansion into spherical harmonics at that value of $ \gamma $ representing the minima in the curves of Figs.~\ref{fig_helix1}(c) and \ref{fig_helix2}(c). Corresponding values are displayed in Figs.~\ref{fig_helix1_spectrum} and \ref{fig_helix2_spectrum} for the configurations of $ r_{helix} = 0.05R $ and $ r_{helix} = 0.1R $, respectively. Again, we plot ten relevant modes for each of the three components of the surface displacement field identified according to the same scheme as in Sec.~\ref{Sec_Global}.


Figure \ref{fig_helix1_spectrum}(a) shows that $ u^{\bot}_{00} $ and $ u^{\bot}_{20} $ dominate the overall behavior (for $ \nu=0.5 $ we correctly find $ u^{\bot}_{00} \approx 0 $). Some higher-order contributions to $ u^{\bot} $ are observed, however, of a relative magnitude of less than 15 \% of the dominant mode, given by either $ u^{\bot}_{00} $ or $ u^{\bot}_{20} $. The configurations are less symmetric than those in Sec.~\ref{Sec_Global}, and we observe a stronger influence of the modes of $ m \neq 0 $, particularly for $ m = 1 $, which characterizes the lowest-order nontrivial dependence on  $ \varphi $.

Next, Fig.~\ref{fig_helix1_spectrum}(b) identifies $ u^{\theta}_{10} $ as a dominating mode of $ u^{\theta} $ for $ \nu \geq 0 $. The same was observed in Fig.~\ref{fig_global_spectrum}(b). In general, higher-order modes enter as well, especially for auxetic materials. As before, the maximal magnitude of the modes described by $ u^{\theta} $ is smaller than the magnitude of the dominating mode for $ u^{\bot} $.

The modes relevant to torsional deformations of the elastic material are addressed in Fig.~\ref{fig_helix1_spectrum}(c). We observe again the most dominant mode to be the lowest one, i.e.\ $ u^{\varphi}_{10} $. However, we also find another mode to be almost equally as strong, namely $ u^{\varphi}_{41} $. This is most likely an effect related to the specific helical structure that was used in our investigation. Nevertheless, both modes are of smaller magnitude when compared to the modes of $ u^{\theta} $ and even smaller when compared to $ u^{\bot} $. Thus, the twisting actuation for this structure is only of secondary importance when compared, for instance, to the global volume change or relative elongation along the magnetization direction.

In addressing the results for the structures of $ r_{helix} =0.1R $ in Fig.~\ref{fig_helix2_spectrum}, we mainly focus on the differences when compared to the situation in Fig.~\ref{fig_helix1_spectrum}. Due to the larger magnitude of $ r_{helix} $, the asymmetry of the configurations with respect to rotations around the $ z $-axis by $ \pi / 3 $ is still more pronounced and we thus observe even stronger modes for $ m \neq 0 $. This trend concerns all three components of the surface displacement field in Figs.~\ref{fig_helix2_spectrum}(a), \ref{fig_helix2_spectrum}(b), and \ref{fig_helix2_spectrum}(c). Differences between Figs.~\ref{fig_helix1_spectrum} and \ref{fig_helix2_spectrum}, especially in the modes for $ m \neq 0 $, can to some extent be traced back to the different value of $ \gamma $ of the investigated structure, according to the different locations of the minima in Figs.~\ref{fig_helix1}(c) and \ref{fig_helix2}(c).

Particularly when focusing on the torsional deformation addressed in Fig.~\ref{fig_helix2_spectrum}(c), we observe that the mode $ u^{\varphi}_{10} $ identifying a global twist deformation is not even the strongest one here. Instead, the strongest mode is $ u^{\varphi}_{31} $. This mode is symmetric for $ z \rightarrow -z $, implying that it cannot describe an overall twist deformation corresponding to a relative rotation of the top hemisphere with respect to the bottom hemisphere. However, the lowest mode of twist actuation $ u^{\varphi}_{10} $ has a much higher absolute magnitude when compared to the structures of $ r_{helix} = 0.05 R $ in Fig.~\ref{fig_helix1_spectrum}(c). Apparently, the radius of the helical elements can have a pronounced effect, partly of antagonistic consequences. If such systems are transferred to actual applications, it is therefore important to adjust the radius of the helical elements to the desired behavior.

Overall, we observe a significantly more pronounced influence of higher-order modes and particularly modes depending on $ \varphi $ for the displacement fields in Figs.~\ref{fig_helix1_spectrum} and \ref{fig_helix2_spectrum} when compared to the results in Fig.~\ref{fig_global_spectrum}. Importantly, the ratio of the magnitudes of $ u^{\varphi}_{10} $ to the magnitudes of $u^{\bot}_{00}$ (except for $ \nu = 0.5 $) or $u^{\bot}_{20}$ is much smaller. Thus, the twist actuation is not as pure for the investigated structures composed of helical elements and we conclude that the globally twisted structures of Sec.~\ref{Sec_Global} are in general more promising candidates to construct a magnetoelastic twist actuator. 
\FloatBarrier

\section{Minimal analytical model} \label{Sec_Force}
\begin{figure}
\centering
\includegraphics[height=4.45cm]{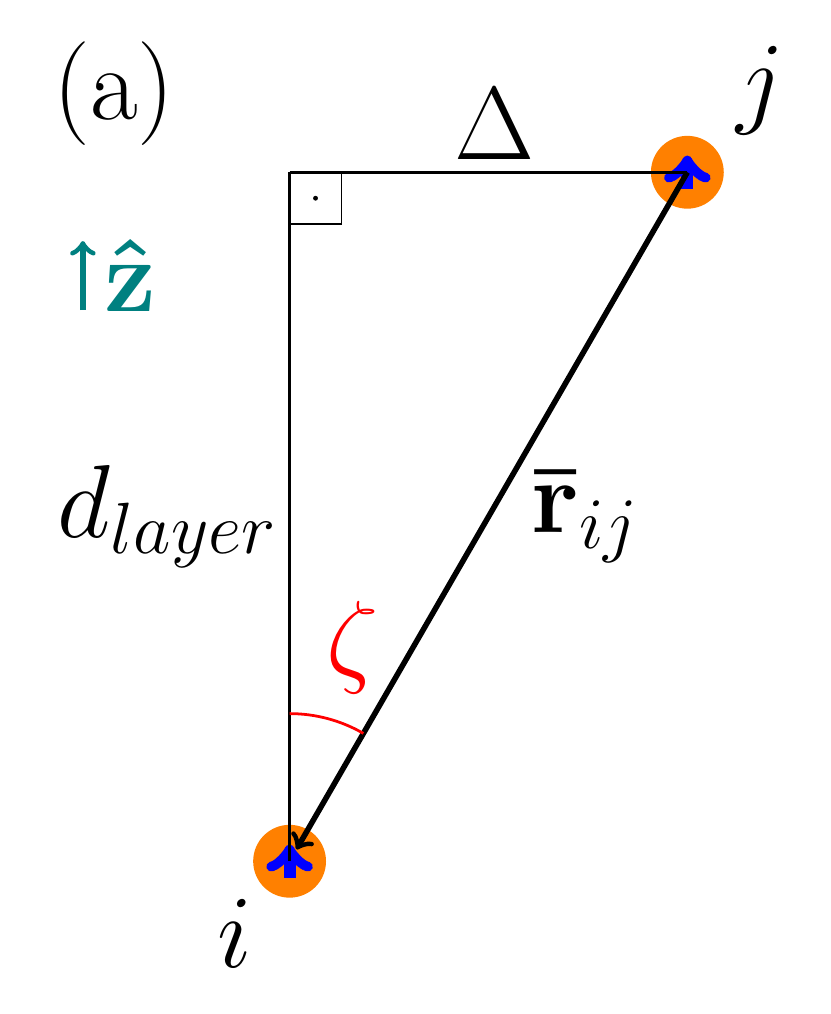}
\includegraphics[height=4.45cm]{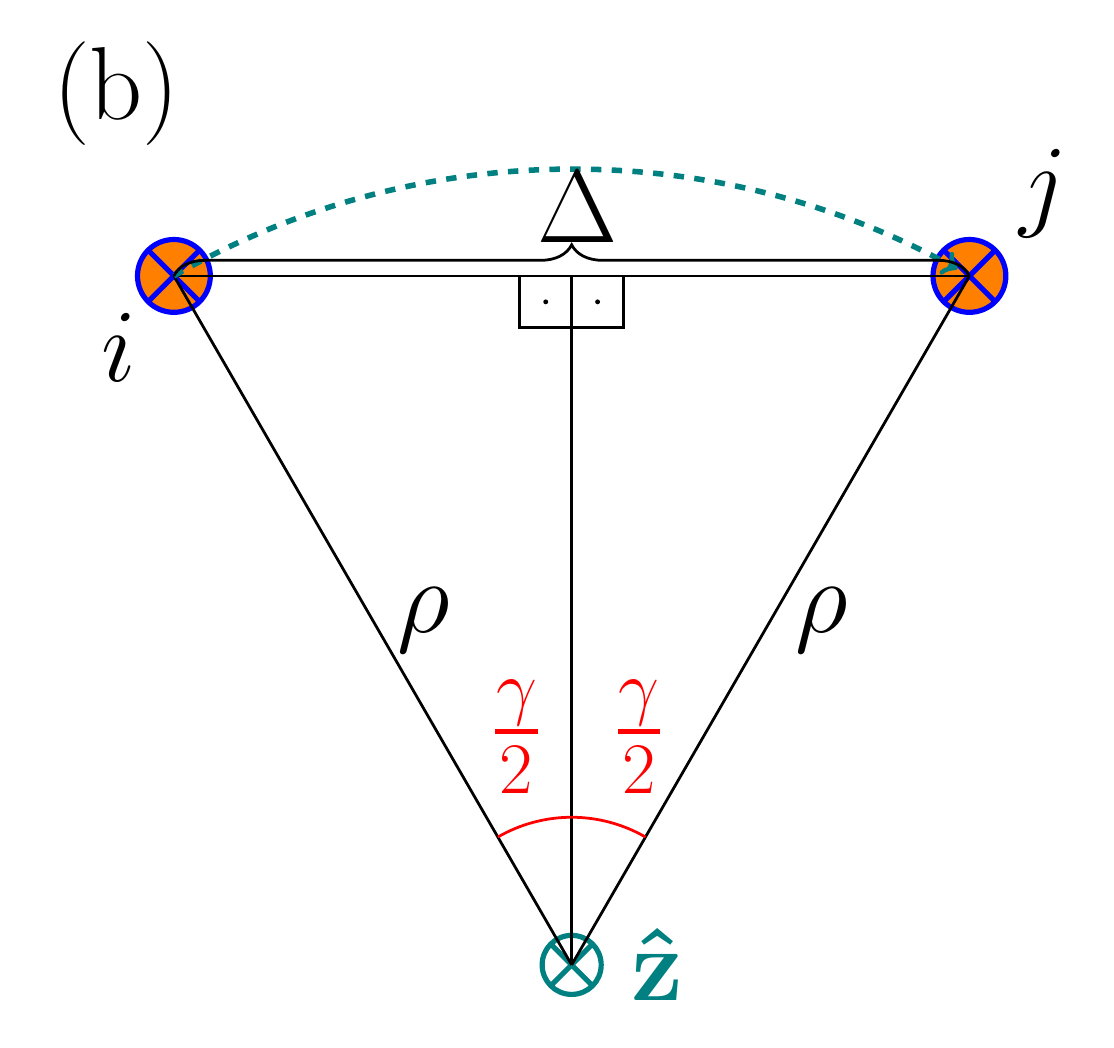}
\caption{In a simplified discussion, we consider the interactions between the magnetized nearest-neighboring particles $ i $ and $ j $ on an initially twisted chain-like aggregate. Their dipole moments, aligned with the center axis $ \mathbf{\hat{z}} $, are depicted by small arrows. We denote the vector from the position of $ j $ to the position of $ i $ by $ \mathbf{\bar{r}}_{ij} $. In (a), their distance along the $ z $-axis is given by $ d_{layer} $ and their lateral distance is denoted as $ \Delta $. $ \zeta $ quantifies the angle between the $ z $-axis and $ \mathbf{\bar{r}}_{ij} $ at the site of particle $ i $. In (b), we show a bottom-view of the configuration. We introduce two right-angled triangles to relate the lateral distance $ \Delta $ between the particles to the radial distance $ \rho $ of the particles from the center axis around which the initial twist of the structure was set. The angle $ \gamma $ was defined previously for both the globally twisted structures and the helically twisted structural elements arranged side by side, see Fig.~\ref{fig_hexagonal} and Eq.~\eqref{eq_helix_vec}, respectively.}
\label{fig_force}
\end{figure}

Having presented our numerical results for the functions $ u^{\varphi}_{10}(\gamma) $ in Secs.~\ref{Sec_Global} and \ref{Sec_Helix}, shown in Fig.~\ref{fig_global}(c) as well as in Figs.~\ref{fig_helix1}(c) and \ref{fig_helix2}(c), respectively, we here discuss how we can understand the behavior qualitatively in simpler terms. To this end, we propose a minimal analytic model based on the dipole--dipole force between the inclusions, see Eq.~\eqref{eq_magn_dipole}. If we only concentrate on the magnetic interactions between two nearest neighbors on a single chain, the geometry can be parameterized as depicted in Fig.~\ref{fig_force}.

Obviously, the situation in reality is more complex as magnetic dipole interactions are long-ranged, leading to magnetic interactions between all particles. Furthermore, due to the magnetically induced deformations the particle positions are affected as well, which changes in turn the magnetic interactions, see Sec.~\ref{Sec_Framework}. Nevertheless, considering pairwise nearest-neighbor interactions along one chain will allow for a basic qualitative description, see below.

Since we are interested in the magnetically induced overall twist deformation, we here focus on the magnetic force components perpendicular to the magnetization direction, i.e.\ in the $ xy $-plane. These in-plane force components are the source of torsional deformations around the $ z $-axis. Instead, the $ z $-components of the magnetic forces are associated with axial contractions. For initially twisted particle configurations and not too large values of $ \gamma $, the in-plane force components represent restoring forces that aim to straighten the chains. Defining $ \zeta $ as the angle between $ \mathbf{\hat{m}} $ and the connecting vector $ \mathbf{\bar{r}}_{ij} $ between two nearest-neighboring particles $ i $ and $ j $, see Fig.~\ref{fig_force}(a), the magnitude $ F_{xy} $ of the in-plane force component on particle $ i $, exerted by particle $ j $, see Eq.~\eqref{eq_magn_dipole}, is given by
\begin{align}\label{eq_force_radial}
F_{xy}(\zeta)  &= \frac{3\mu_0 m^2}{4\pi} \frac{\cos^4\zeta}{d_{layer}^4}
			  	\left| 5  \sin\zeta \cos^2\zeta - \sin\zeta \right|.
\end{align}
Here, we have inserted $ \mathbf{\hat{m}} \cdot \mathbf{\hat{\bar{r}}}_{ij} = \cos \zeta $, $ \bar{r}_{ij} = d_{layer} / \cos \zeta $, and $ \sin \zeta $ for the component of $ \mathbf{\hat{\bar{r}}}_{ij} $ in the $ xy $-plane.

Next, we maximize $ F_{xy}(\zeta) $ with respect to $ \zeta $ to find out which configuration of particles $ i $ and $ j $ leads to a maximized restoring force, which supports a maximized twist actuation. The maximum is found for
\begin{align}
\cos^2\zeta_{max} = \frac{1}{2} + \frac{1}{2} \sqrt{\frac{19}{35}}.
\end{align}
If we now restrict the solutions to the range $ 0 < \zeta < \pi / 2 $, we find the unique solution
\begin{align} \label{eq_zeta_max}
\zeta_{max} = \arccos\left(\sqrt{ \frac{1}{2} + \frac{1}{2} \sqrt{\frac{19}{35}} }\right)
\approx 0.118 \pi.
\end{align}
When we compare to our previous results, we can use the relations deduced from Fig.~\ref{fig_force}(a)
\begin{align} \label{eq_tan_zeta}
\tan\zeta = \frac{\Delta}{d_{layer}}
\end{align}
and Fig.~\ref{fig_force}(b)
\begin{align} \label{eq_gamma}
\sin\left(\frac{\gamma}{2} \right) = \frac{\Delta}{2\rho},
\end{align}
where we have introduced $ \rho $ as the distance of the inclusions $ i $ and $ j $ from the axis of the initial twist of the corresponding structure. To relate the result of this analytical consideration to our numerical evaluation, we find from Eqs.~\eqref{eq_tan_zeta} and \eqref{eq_gamma}
\begin{align} \label{eq_gamma_max}
\gamma_{max} = 2 \arcsin\left( \frac{d_{layer}}{2\rho} \tan\zeta_{max} \right)
\end{align} 
where $ \gamma_{max} $ implies a maximized in-plane torsional force component, based on this simplified analytical consideration.
For the systems addressed in Sec.~\ref{Sec_Helix}, to compare these analytical and the numerical results, we can simply set $ \rho = r_{helix} $. For the globally twisted configurations in Sec.~\ref{Sec_Global}, the situation is more complex as there is not a single value of $ \rho $ that is equal for all chain-like aggregates, but the value of $ \rho $ depends on which chain we consider.

\begin{figure}
\centering
\includegraphics[width=7.64468cm]{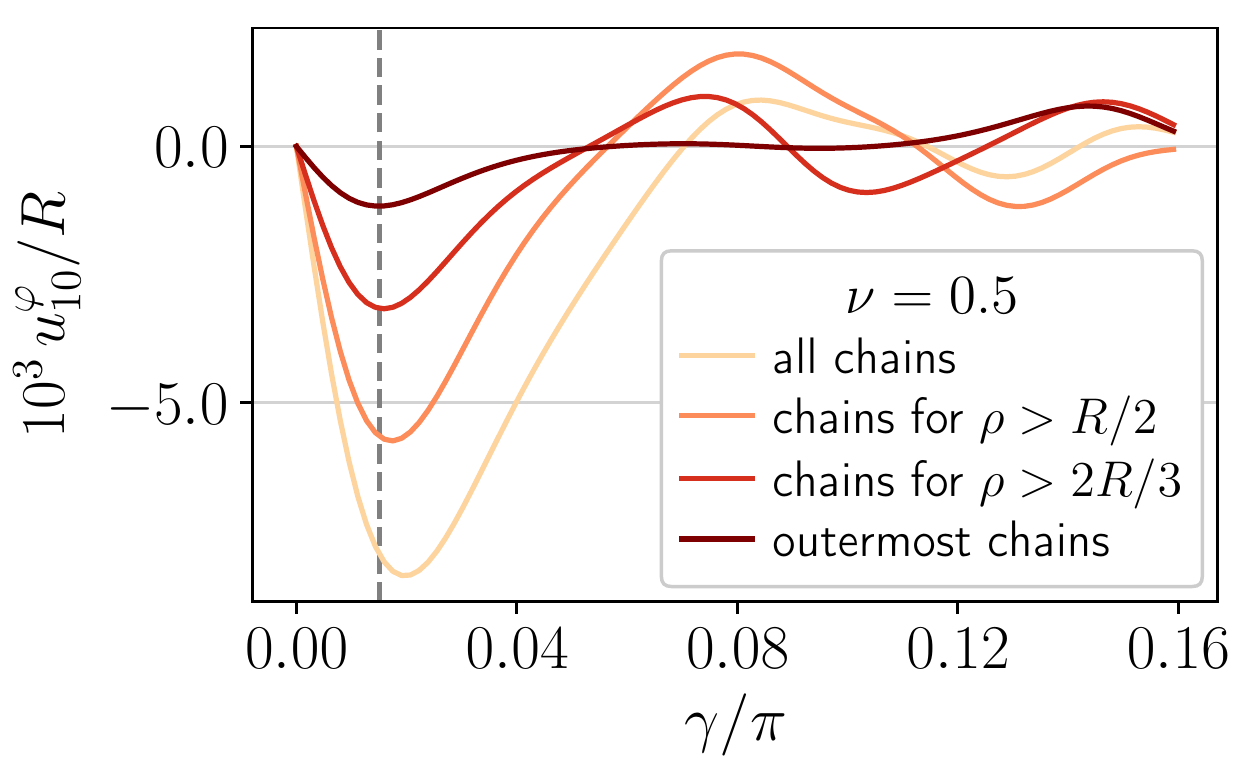}
\caption{Same as in Fig.~\ref{fig_global}(c), but for configurations for which we only consider those chains that have a minimal distance $ \rho $ from the axis of twist of the initial nonmagnetized structure. We show a comparison between the results of Fig.~\ref{fig_global}(c), here labeled as ``all chains'', and corresponding configurations that only include those chains for which $ \rho > R/2 $ and $ \rho > 2R/3 $. Furthermore, we show results for only keeping the outermost chains, i.e.\ chains of $ \rho = \sqrt{13} \, d_{chain} \approx 0.901 R $. In all cases, we only display the results for incompressible systems, i.e.\ for $ \nu = 0.5 $, for clarity. Particularly, we note how the position of the global minimum is slightly shifted towards smaller values of $ \gamma $ for configurations of larger average values of $ \rho $ for the chains. The vertical gray dashed line marks the value $ \gamma_{max} \approx 0.015 \pi $ as obtained from Eqs.~\eqref{eq_zeta_max} and \eqref{eq_gamma_max} for the outermost chains.}
\label{fig_global_compare}
\end{figure}

To illustrate this more complex dependence for the globally twisted structures on the angle $ \gamma $, quantified by $ u^{\varphi}_{10} (\gamma)$, we have generated additional globally twisted configurations while removing from the systems considered in Sec.~\ref{Sec_Global} those chain-like elements that have a value of $ \rho $ smaller than a certain threshold. Illustratively, this corresponds to only considering those chains that are located outside a coaxial cylinder of diameter $ 2\rho $. In Fig.~\ref{fig_global_compare}, we present results for cut-off values of $ \rho $ of $ R/2 $, $ 2R/3 $, and $ \sqrt{13} \, d_{chain} \approx 0.901 R$, where the latter value marks the outermost chains. For comparison, we have added in Fig.~\ref{fig_global_compare} the results for the configurations of Sec.~\ref{Sec_Global} as well.  For this evaluation, we  restrict ourselves to incompressible elastic materials ($ \nu=0.5 $) for clarity.

The main result of Fig.~\ref{fig_global_compare} is that as we increase the lower threshold value of $ \rho $, the global minimum is shifted towards lower values of $ \gamma $.
For all chains considered, see Sec.~\ref{Sec_Global}, the value of $ \gamma $ corresponding to a maximized twist deformation is $ \gamma \approx 0.019 \pi $. Introducing a cut-off for $ \rho $ of $ R/2 $, this value is reduced to $ \gamma \approx 0.018 \pi $. Moving on to a cut-off for $ \rho $ of $ 2R/3 $, it is further reduced to  $ \gamma \approx 0.016 \pi $. When keeping only the outermost chains, we obtain  $ \gamma \approx 0.014 \pi $ for the location of the maximized twist deformation.
Moreover, we observe a decrease in magnitude of the minimum of $ u^{\varphi}_{10} $. This contains, however, a trivial effect as we decrease the number of inclusions for increasing cut-off values for $ \rho $. More precisely, we find 623, 324, 168, and 60 inclusions for the four different systems addressed in Fig.~\ref{fig_global_compare}. 

When we now compare our numerical results to the minimal analytical model according to Eqs.~\eqref{eq_zeta_max} and \eqref{eq_gamma_max}, we consider the configurations of only keeping the outermost chains. In this case, inserting $ \rho \approx 0.901 R $ into Eq.~\eqref{eq_gamma_max}, we obtain a value of $ \gamma_{max} \approx 0.015 \pi $, see the vertical dashed line in Fig.~\ref{fig_global_compare}. This is only slightly bigger than the numerical value of $ \gamma \approx 0.014 \pi $. 
It shows a fair agreement, considering for instance the assumptions of including only nearest-neighbor particle interactions and rigid particle positions in the analytical model.

Next, we compare the numerical and analytical results for the structures composed of helical elements as considered in Sec.~\ref{Sec_Helix}. Setting $ \rho = r_{helix} $, we find from the analytical consideration $ \gamma_{max} \approx 0.28 \pi $ and $ \gamma_{max} \approx 0.14 \pi $ for $ r_{helix} = 0.05 R $ and $ r_{helix} = 0.1 R $, respectively. The results of our numerical investigation for $ u^{\varphi}_{10} $ were $ \gamma \approx 0.24 \pi $ and $ \gamma \approx 0.13 \pi $, respectively, see Sec.~\ref{Sec_Helix}. While showing fair agreement concerning the involved approximations, our analytical model again shows a tendency of overestimating the numerical results, see above.

Within our minimal analytical model, we may equally well estimate analytically the lowest value of $ \gamma > 0$ for which $ u^{\varphi}_{10} $ becomes zero. Again, we require a fixed value of $ \rho $.
From Eq.~\eqref{eq_force_radial}, we find that $ F_{xy} = 0 $ for a value $ \zeta_{0} > 0 $ of
\begin{align}
\zeta_{0} = \arccos\left(\frac{1}{\sqrt{5}} \right) \approx 0.352\pi,
\end{align}
corresponding to
\begin{align}
\gamma_{0} = 2 \arcsin\left( \frac{d_{layer}}{2\rho} \tan\zeta_{0} \right) 
= 2 \arcsin\left( \frac{d_{layer}}{\rho} \right).
\end{align}
Inserting the value of $ \rho $ for the outermost chains in Fig.~\ref{fig_global_compare} implies $ 
\gamma_{0} \approx 0.078 \pi $, while the numerical result for $ u^{\varphi}_{10} $ suggests $ \gamma \approx 0.057 \pi $. As before, we observe that the analytically determined value of $ \gamma $ exceeds the one determined numerically. For the systems containing the helical structural elements, our analytical estimate does not imply any value of $ 0 < \gamma < \pi $ at which $ F_{xy}=0 $ because $ d_{layer} > r_{helix} $ in both cases. This is in line with our numerical results for $ r_{helix} = 0.05 R $, for which $ u^{\varphi}_{10} < 0 $ for all $ 0 < \gamma < \pi $. However, our numerical investigation reveals a value of $ \gamma \approx 0.62 \pi $ at which $ u^{\varphi}_{10} = 0 $ for $ r_{helix} = 0.1 R $.

\begin{figure}
	\includegraphics[width=\linewidth]{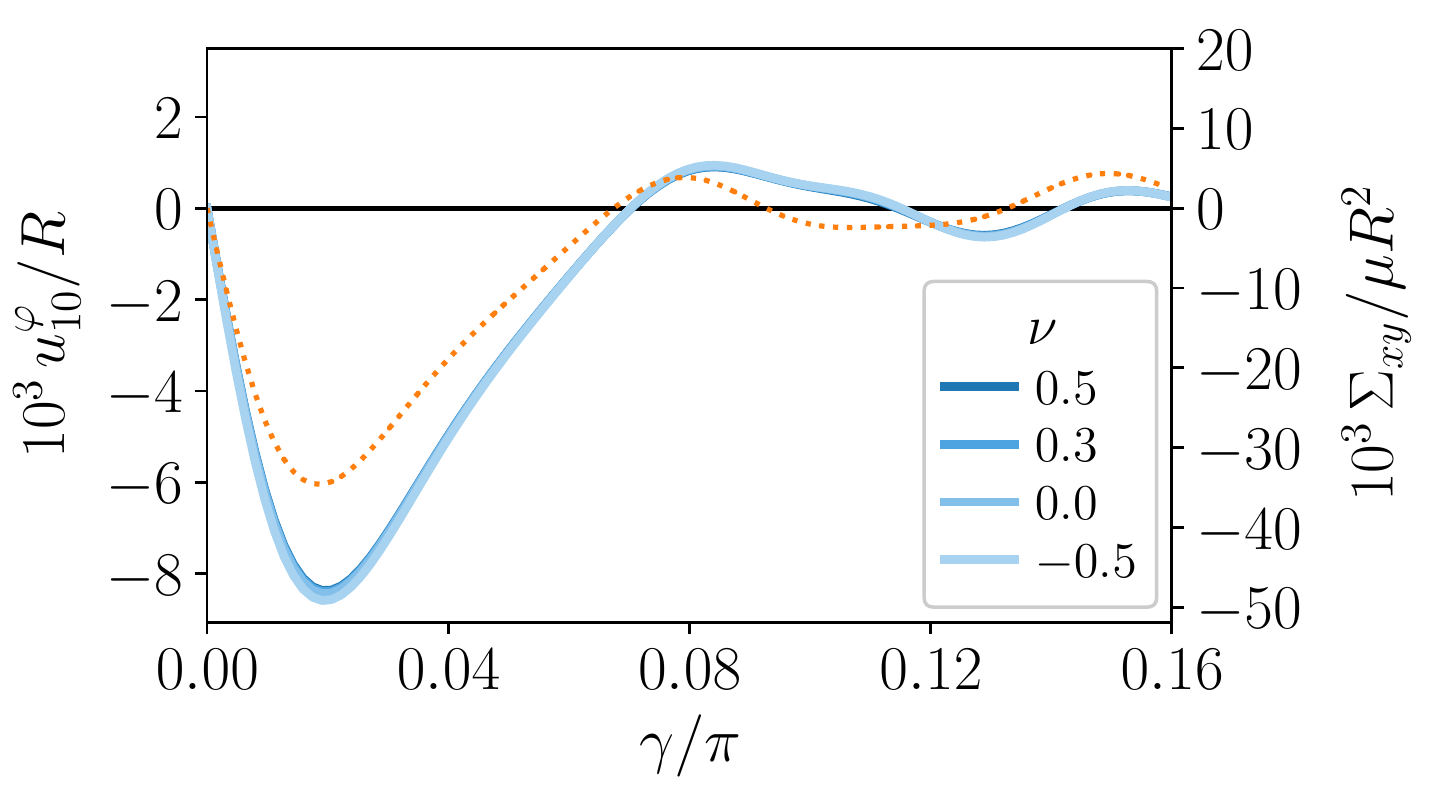}
	\caption{Sum $\Sigma_{xy}$ over the appropriately signed azimuthal magnetic force components acting on all particles as defined in the main text (dotted line) for the systems containing the globally twisted structures. The shape of the curve for $\Sigma_{xy}$ qualitatively agrees with the shape of the curves for $u_{10}^{\varphi}$ reproduced from Fig.~\ref{fig_global}(c) (solid lines) that quantify the induced overall torsional deformation.}
	\label{fig_global_force}
\end{figure}

As had become obvious above and from Fig.~\ref{fig_global_compare}, comparing the simple analytical model approach to the numerical results for the complete globally twisted structures is less direct. The different chain-like aggregates in the system are located at different radial distances $\rho$ from the center axis. These varying distances need to be taken into account.
	
To find a reasonable measure, we start from the magnetic forces $\mathbf{F}_i$ according to Eq.~\eqref{eq_magn_dipole} on each particle $i$. We denote by $\hat{\boldsymbol{\varphi}}_i$ the local azimuthal unit vector in the spherical coordinate system at the position of particle $i$. To identify those force components that supposedly directly support the overall twist deformation, we project $\mathbf{F}_i$ onto $\hat{\boldsymbol{\varphi}}_i$ on the upper hemisphere and onto $-\hat{\boldsymbol{\varphi}}_i$ on the lower hemisphere. Particles $ i $ located on the center axis and on the equatorial plane are not taken into account. Finally, the sum over all force components obtained in this way is denoted as $ \Sigma_{xy} $. It is plotted as the dotted line in Fig.~\ref{fig_global_force}, together with the results for $u_{10}^{\varphi}$ as displayed in Fig.~\ref{fig_global}(c). Since the initial positions of the particles are used for this basic analytical evaluation, the curves for $ \Sigma_{xy} $ are independent of the value of the Poisson ratio $\nu$.

Comparing these graphs, we notice that the force component $ \Sigma_{xy} $ as well as the curves for $ u^{\varphi}_{10} $ have a pronounced minimum at approximately the same value of $ \gamma \approx 0.019 \pi $. Moreover, for the lowest value of $ \gamma > 0 $ at which $ \Sigma_{xy} = 0 $ and $ u^{\varphi}_{10}=0 $, we find $ \gamma \approx 0.068 \pi $ and $ \gamma \approx 0.072 \pi $, respectively.

In summary, we can estimate certain characteristic points on the curves of $ u^{\varphi}_{10}(\gamma) $ by simple analytical model considerations. Often, it is sufficient to focus on the interactions between neighboring dipoles only. For the globally twisted structures, see Sec.~\ref{Sec_Global}, the different distances of the chain-like aggregates from the center axis of the elastic sphere need to be taken into account for more quantitative evaluations. Nonaffine elastic deformations have not been included in the simple analytical model. Furthermore, our simplified analytical approach does not account for the change in magnetized particle positions during deformations included in our numerical description.
\FloatBarrier

\section{Conclusions}\label{Sec_final}
To conclude, we have suggested a way to construct soft torsional actuators using magnetic gels and elastomers. For this purpose, we have addressed two different structural arrangements of the magnetizable inclusions in the elastic material: globally twisted structures and side-by-side arrangements of helical elements. Both are generated from initially hexagonally arranged parallel chain-like elements. For both configurations, we have explicitly calculated the resulting magnetostrictive distortion of the overall system upon magnetization. In this context, for reasons of analytical accessibility, we have here concentrated on systems of overall spherical shape. Particularly, we have focused on the degree of induced twist actuation, which we quantified using a spherical harmonic mode of expansion of the surface displacement field.

Among the systems that we investigated, we found the globally twisted structures to show a significantly larger twist actuation when compared to the systems containing helical elements arranged side by side. For the studied globally twisted structures, the overall deformational response is indeed dominated by a twist-type distortion. Instead, the overall twist response in the case of the embedded helical elements arranged side by side was less pure. Thus, it appears that the considered globally twisted structures are better suited to construct a soft torsional actuator. In the near future, these might also be the ones requiring less additional effort for actual fabrication.

Furthermore, we have quantified which degree of initial structural twist in the nonmagnetized state leads to a maximized twist actuation. Such an optimized value arises from two antagonistic tendencies. If the internal structure is not twisted at all, then an overall torsional deformation cannot be induced. However, if the initial structure is twisted too much, then the interactions between the inclusions upon magnetization even become repulsive for a too large lateral separation of the contained magnetizable particles. We find an optimized value in between. In fact, these properties can be understood already on a qualitative basis by addressing the magnetic interactions between two neighboring particles on one initially deformed chain-like structural aggregate.

Overall, we hope that our study will inspire experimental realizations in the future. 
Corresponding devices may find possible applications, for instance, as microfluidic mixing actuators. Not only can twist deformations and thus torsional flows around such an element be induced upon request from outside by alternating magnetic fields. But also, due to the existence of an overall structural anisotropy identified, for example, by the axis of global twist, can the mixing element simultaneously be oriented by the direction of the external field.
Moreover, as long as dynamic effects like leaking electrical currents do not play an important role, our results equally apply to the construction of corresponding devices from electrorheological gels and elastomers \cite{an2003actuating,allahyarov2015simulation,liu2001electrorheology}, using external electric fields for actuation.

\begin{acknowledgments}
Some of the results in this paper have been derived using the
HEALPix package \cite{HEALPix}.
The authors thank the Deutsche Forschungsgemeinschaft for support of this work through the priority program SPP 1681, grant no.~ME 3571/3.
\end{acknowledgments}


\bibliography{lit_fischer}
\end{document}